\documentclass[twocolumn,trackchanges,floatfix]{aastex631}
\usepackage{amsmath}
\usepackage{longtable}
\usepackage{multirow}
\usepackage{booktabs}

\newcommand{\AGAMA}{\texttt{AGAMA}}
\newcommand{\MWMMXVII}{\texttt{MW2017}}
\newcommand{\HDBSCAN}{\texttt{HDBSCAN}}
\newcommand{\peri}{\text{peri}}
\newcommand{\apo}{\text{apo}}
\newcommand{\maxtext}{\text{max}}

\renewcommand{\arraystretch}{0.5}

\shorttitle{RAVE DTGs}
\shortauthors{Shank et al.}

\begin{document}

\title{Dynamically Tagged Groups of Metal-Poor Stars II. The Radial Velocity Experiment Data Release 6}

\author[0000-0001-9723-6121]{Derek Shank}
\affiliation{Department of Physics, University of Notre Dame, Notre Dame, IN 46556, USA}
\affiliation{Joint Institute for Nuclear Astrophysics -- Center for the Evolution of the Elements (JINA-CEE), USA}

\author[0000-0003-2159-4760]{Dante Komater}
\affiliation{Department of Physics, University of Notre Dame, Notre Dame, IN 46556, USA}

\author[0000-0003-4573-6233]{Timothy C. Beers}
\affiliation{Department of Physics, University of Notre Dame, Notre Dame, IN 46556, USA}
\affiliation{Joint Institute for Nuclear Astrophysics -- Center for the Evolution of the Elements (JINA-CEE), USA}

\author[0000-0003-4479-1265]{Vinicius M. Placco}
\affiliation{NSF’s NOIRLab, 950 N. Cherry Ave., Tucson, AZ 85719, USA}

\author[0000-0003-3250-2876]{Yang Huang}
\affiliation{South-Western Institute for Astronomy Research, Yunnan University, Kunming 650500, People's Republic of China}

\date{\today}

\begin{abstract}
    Orbital characteristics based on Gaia Early Data Release 3 astrometric parameters are analyzed for ${\sim} 8,000$ metal-poor stars ([Fe/H] $\leq -0.8$) compiled from the RAdial Velocity Experiment (RAVE) Data Release 6. Selected as metal-poor candidates based on broadband photometry, RAVE collected moderate-resolution ($R \sim 7,500$) spectra in the region of the Ca triplet for these stars. About $20\%$ of the stars in this sample also have medium-resolution ($1,200 \lesssim R \lesssim 2,000$) validation spectra obtained over a four-year campaign from $2014$ to $2017$ with a variety of telescopes. We match the candidate stars to photometric metallicity determinations from the Huang et al. recalibration of the Sky Mapper Southern Survey Data Release 2. We obtain dynamical clusters of these stars from the orbital energy and cylindrical actions using the \HDBSCAN ~unsupervised learning algorithm. We identify $179$ Dynamically Tagged Groups (DTGs) with between $5$ and $35$ members; $67$ DTGs have at least $10$ member stars. Milky Way (MW) substructures such as Gaia-Sausage-Enceladus, the Metal-Weak Thick Disk, the Splashed Disk, Thamnos, the Helmi Stream, and LMS-1 (Wukong) are identified. Associations with MW globular clusters are determined for $10$ DTGs; no recognized MW dwarf galaxies were associated with any of our DTGs. Previously identified dynamical groups are also associated with our DTGs, with emphasis placed on their structural determination and possible new identifications. We identify chemically peculiar stars as members of several DTGs; we find $22$ DTGs that are associated with \textit{r}-process-enhanced stars. Carbon-enhanced metal-poor (CEMP) stars are identified among the targets with available spectroscopy, and we assign these to morphological groups following the approach given by Yoon et al.
    
\end{abstract}

\keywords{Milky Way dynamics (1051), Galaxy dynamics (591), Galactic archaeology (2178), Milky Way evolution (1052), Milky Way formation (1053), Milky Way stellar halo (1060)}

\section{Introduction}\label{sec:Introduction}

Large-scale spectroscopic surveys conducted over the last few decades have allowed the structure, assembly, and chemical-evolution history of the Milky Way (MW) to be explored in great detail \citep{York2000,Yanny2009,Cui2012}. The spectra from such surveys provide important information about stellar atmospheres, in particular their chemical abundances, an important tool for determining the origin and evolution of the elements in the stellar populations of the Galaxy. The collected spectra can also provide radial velocity measurements, which are used as one of the $6$-D astrometric parameters (along with position, distance, and proper motions) to derive the orbits of stars in a selected gravitational potential. 

The Radial Velocity Experiment (RAVE; \citealt{Steinmetz2006}) is one of the largest contributors of such information.  In the newest RAVE Data Release 6 (DR6; \citealt{Steinmetz2020a}) there are $\sim 450,000$ unique stellar objects with radial velocities available, with $95\%$ having an accuracy better than $4.0$ km s$^{-1}$. With such a large and homogeneous set of radial velocities available, RAVE provides the opportunity to study the MW through both their dynamics and chemical compositions. One of the first uses of RAVE to study MW dynamics was by \citet{Seabroke2008}, who determined that there were no vertical tidal streams associated within the Solar Neighborhood, the region surveyed by RAVE Data Release 1 (DR1; \citealt{Steinmetz2006}). Also using RAVE DR1, \citet{Klement2008} discovered a new radial stream, while recovering previously known streams such as the Helmi Stream \citep{Helmi1999}. The stellar parameters obtained by RAVE have allowed astronomers to discover metal-poor stars ([Fe/H]\footnote{The standard definition for an abundance ratio of an element in a star $(\star)$ compared to the Sun $(\odot)$ is given by $[A/B] = (\log{N_{A}/N_{B}})_{\star} - (\log{N_{A}/N_{B}})_{\odot}$, where $N_{A}$ and $N_{B}$ are the number densities of atoms for elements $A$ and $B$.} $\lesssim -1.0$), as first reported by \citet{Fulbright2010} using both DR1 and Data Release 2 (DR2; \citealt{Zwitter2008}).  

\citet{Cocskunovglu2011,Cocskunovglu2012}, \citet{Bilir2012}, \citet{Duran2013}, and \citet{Karaali2014} employed kinematics and stellar parameters from RAVE to determine estimates of the Local Standard of Rest, radial and vertical metallicity gradients, and space velocity components for the thin and thick disk of the Galaxy. \citet{Antoja2012} was one of the first to probe beyond the Solar Neighborhood in search of kinematic groups using RAVE. These authors reidentified some known groups, such as Hercules, while also recovering new kinematic over-densities, revealing non-axisymmetric groups present in the MW. Utilizing the full RAVE Data Release 4 (DR4; \citealt{Kordopatis2013}), \citet{Binney2014} was able to compare the kinematics of stars within $2$ kpc of the Sun to dynamical models, showing remarkable consistency for the velocity components of the sample.

The advent of Gaia \citep{GaiaCollaboration2016b} has provided proper motions and parallax-based distances for an unprecedented number of stars. Using both Gaia Data Release 1 (DR1; \citealt{GaiaCollaboration2016a}) and RAVE DR4, \citet{Helmi2017} argued that the stellar halo was built solely by mergers and had a dominant retrograde-velocity component. Using the same data sets, \citet{Robin2017} were able to constrain the formation histories of the thin and thick disks. Recently, \citet{Li2020} employed RAVE Data Release 5 (DR5; \citealt{Kunder2017}) and Gaia Data Release 2 (DR2; \citealt{GaiaCollaboration2018}) to discover new structures in the MW using dynamical quantities such as the orbital energy and angular momentum, which offer insight into the structures' origins. 

When Galactic satellites are accreted and dispersed into the MW, the energies and dynamical actions of their member stars are expected to resemble those of their parent progenitor satellites \citep{Helmi1999}. The seminal work of \citet{Roederer2018a} employed unsupervised clustering algorithms, an approach that has proven crucial to determine structures in the MW that are not revealed through large-scale statistical sampling methods. These authors were able to collect $35$ chemically peculiar (\textit{r}-process-enhanced) stars and determine their orbits. With these data in hand, multiple clustering tools were applied to the orbital energy and actions to determine stars with similar orbital characteristics. This study revealed eight dynamical groupings comprising between two and four stars each. The small dispersion of each group's metallicity was noted, and accounted for by reasoning that each group was associated with a unique satellite accretion event.

\citet{Yuan2020b} utilized the self-organizing map neural network routine StarGO \citep{Yuan2018} on the Large Sky Area Multi-Object Fiber Spectroscopic Telescope (LAMOST; \citealt{Cui2012}) Data Release 3 \citep{Li2018b} stellar survey. These authors used StarGO to examine the very metal-poor ([Fe/H] $\lesssim -1.8$) stars to seek dynamical clusters based on the derived energy, angular momentum, and polar and azimuthal (E,L,$\theta$,$\phi$) parameters. From this prescription, the authors identified $57$ dynamically tagged groups (DTGs), of which most were associated with GSE or Sequoia \citep{Myeong2019}, while $18$ were new structures not previously associated with known large-scale substructures. \citet{Limberg2021a} constructed DTGs from metal-poor stars in the HK \citep{Beers1985,Beers1992} and Hamburg/ESO \citep{Christlieb2008} surveys using the Hierarchical Density-Based Spatial Clustering of Applications with Noise (\HDBSCAN; \citealt{Campello2013}) algorithm over the orbital energy and cylindrical action space. Their clustering procedure was able to identify 38 DTGs, with 10 of those being newly identified substructures. \citet{Gudin2021} extended the work by \citet{Roederer2018a}, using a much larger sample of \textit{r}-process-enhanced stars (see their Table 1 for definitions). Also utilizing the \HDBSCAN ~algorithm, 30 Chemo-Dynamically Tagged Groups (CDTGs)\footnote{The distinction between CDTGS and DTGs is that the original stellar candidates of CDTGs are selected to be chemically peculiar in some fashion, while DTGs are selected from stars without detailed knowledge of their chemistry, other than [Fe/H].} were discovered. Their analysis revealed statistically significant similarities in the dispersions of stellar metallicity, carbon abundance, and \textit{r}-process-element ([Sr/Fe], [Ba/Fe], and [Eu/Fe]) abundances, strongly suggesting that these stars experienced similar chemical-evolution histories in their progenitor galaxies.

\begin{figure*}[t]
    \includegraphics[width=\textwidth]{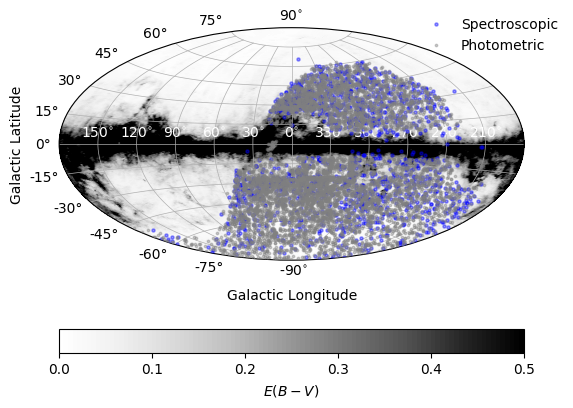}
    \caption{The Galactic positions of the Full Sample of RAVE DR6 stars with the spectroscopic subset shown as blue points and the photometric subset as gray points. The Galactic reddening map, taken from \citet{Schlegel1998}, and recalibrated by \citet{Schlafly2011}, is shown in the background on a gray scale with darker regions corresponding to larger reddening.}
    \label{fig:galactic_map}
\end{figure*}

This work aims to analyze the DTGs present among stars in RAVE Data Release 6 (DR6; \citealt{Steinmetz2020a}), focusing on metal-poor ([Fe/H] $\leq -0.8$) stars. The procedures employed closely follow the work of \citet{Shank2021}, hereafter referred to as Paper I in this series, which considered DTGs found in the sample of the Best \& Brightest selection of \citet{Schlaufman2014}. The association of our identified DTGs with recognized Galactic substructures, previously known DTGs/CDTGs, globular clusters, and dwarf galaxies is explored, with the most interesting stellar populations being noted for future high-resolution follow-up studies.

This paper is outlined as follows. Section \ref{sec:Data} describes the RAVE DR6 sample, along with their associated astrometric parameters and the dynamical parameters. The clustering procedure is outlined in Section \ref{sec:ClusteringProcedure}. Section \ref{sec:StructureAssociations} explores the clusters and their association to known MW structures. 
Finally, Section \ref{sec:Discussion} presents a short discussion and perspectives on future directions.

\section{Data}\label{sec:Data}

The RAVE DR6 survey \citep{Steinmetz2020a} forms the basis for the compilation of our data set.
We construct three samples, the Full Sample, the Initial Sample, and the Final Sample, which are described in the following sections.

\subsection{Construction of the Full Sample}

\begin{figure*}[t]
    \includegraphics[width=\textwidth]{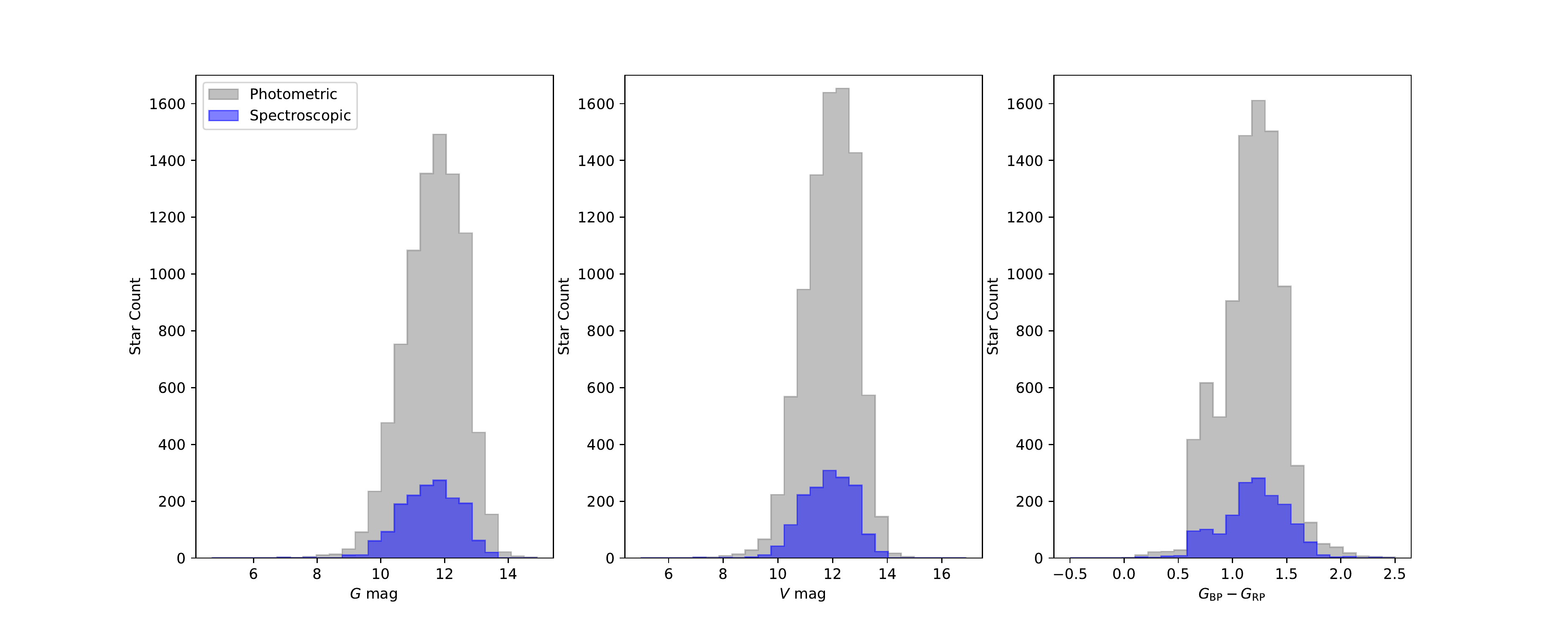}
    \caption{Left Panel: Histogram of the $G_{\text{mag}}$ for the Full Sample. Middle Panel: Histogram of the $V_{\text{mag}}$ for the Full Sample. Right Panel: Histogram of the $G_{\text{BP}} - G_{\text{RP}}$ color for the Full Sample. All Panels: The spectroscopic subset of the Full Sample is represented as a blue histogram; the photometric subset is represented as a gray histogram.}
    \label{fig:mag_hist}
\end{figure*}

Of the $322,367$ unique stars in RAVE DR6 that have acceptable quality-flags from the \texttt{MADERA} stellar parameter pipeline \citep{Steinmetz2020b}, photometric metallicity and temperature estimates are taken from the SkyMapper Southern Survey (SMSS; \citealt{Wolf2018}) Data Release 2 (DR2; \citealt{Onken2019}). These estimates are derived by the procedure described in \citet{Huang2021c} and explained below. Metal-poor candidate stars from RAVE are also taken from the sample in \citet{Placco2018} whose authors explored the nature of the RAVE stellar parameters with a corresponding set of spectroscopic parameters, also explained below.

The validation spectra from \citet{Placco2018}\footnote{A comprehensive explanation of the construction of the spectroscopic sample, along with details of the medium-resolution spectroscopic observations, are discussed in \citet{Placco2018} and Paper I, to which we refer the interested reader.} were then used to determine the stellar atmospheric parameters and elemental abundances for the stars using the non-SEGUE Stellar Parameter Pipeline (n-SSPP; \citealt{Beers2014,Beers2017}), following the procedures described in Paper I. The  parameters obtained are the effective temperature (T$_{\rm eff~Spec}$), surface gravity (log \textit{g}), and metallicity ([Fe/H]$_{Spec}$), while the elemental abundances are the carbon abundance ([C/Fe]), and the $\alpha$-element abundance ([$\alpha$/Fe]).  The carbon abundance is then adjusted, using the prescription outlined in \citet{Placco2014}, to account for the depletion of carbon along the red giant branch. This corrected carbon abundance ([C/Fe]$_{c}$) is used as the star's natal carbon abundance. The average errors adopted for each of the stellar parameters for the spectra with S/N $\sim 30$ are $\pm 150$ K for $T_{\text{eff}}$; $\pm 0.35$\,dex for log \textit{g}, and $\pm 0.20$\,dex for [Fe/H], [C/Fe], and [$\alpha$/Fe], with values for each star listed in Table~\ref{tab:initial_data_descript} in the Appendix (See \citealt{Lee2008a} for more information on the errors).  Note that the elemental abundances reported here supersede those published in \citet{Placco2018}\footnote{The [C/Fe] and $\alpha$-element abundances present in \citet{Placco2018} were found to be systematically offset; the updated values are listed here. The stellar parameters were not affected, in general, although there are occasionally different reported [Fe/H]$_{Spec}$ values reported based on later re-inspections of the spectra by Beers.}.

\begin{figure*}[t]
    \includegraphics[width=\textwidth]{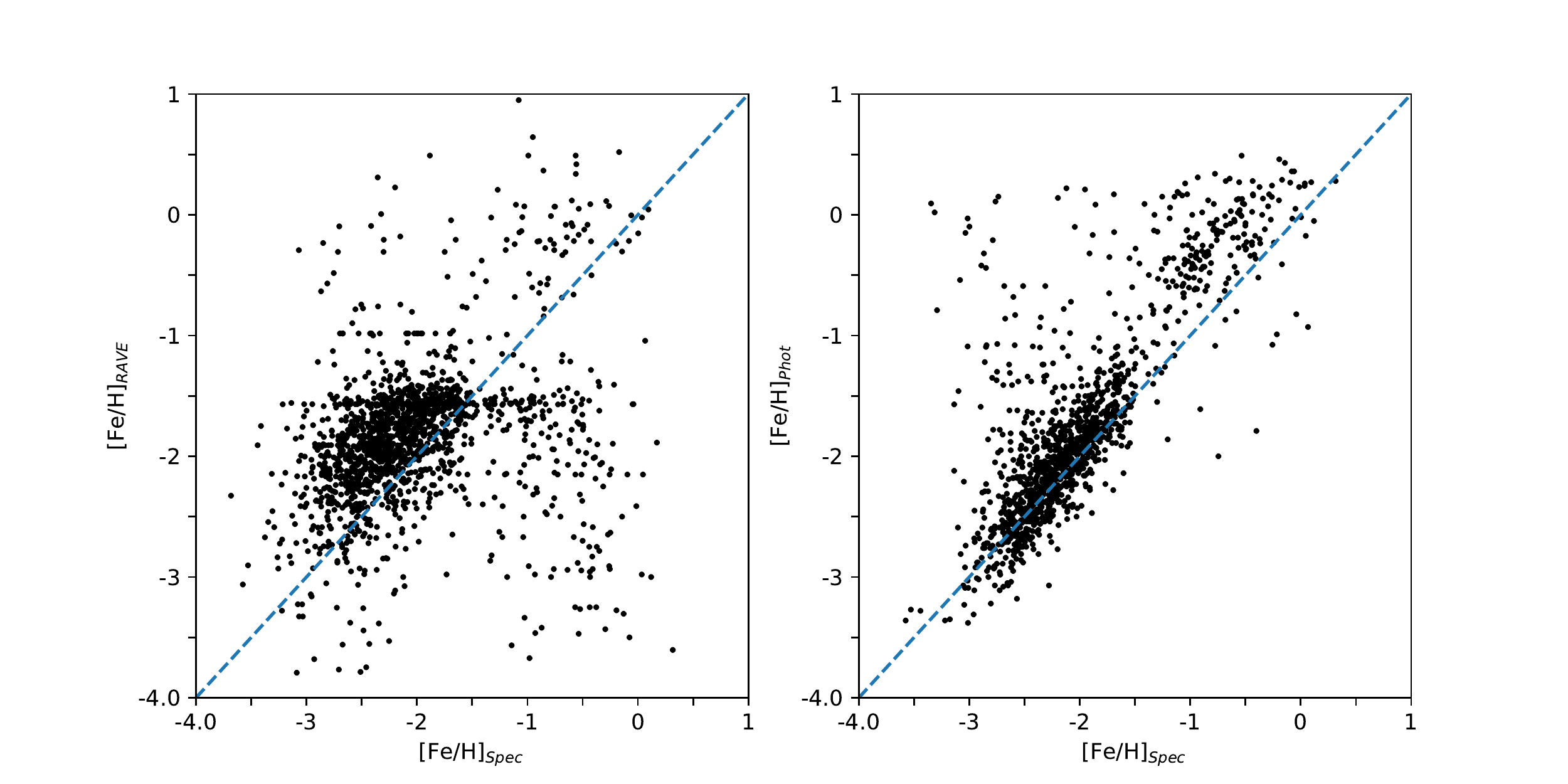}
    \caption{Left: Comparison plot between the spectroscopic metallicity from the n-SSPP ([Fe/H]$_{Spec}$) and the RAVE DR6 metallicity ([Fe/H]$_{RAVE}$) in the Full Sample. Right: Comparison plot between the spectroscopic metallicity from the n-SSPP ([Fe/H]$_{Spec}$) and the photometric metallicity estimates from \cite{Huang2021c} ([Fe/H]$_{Phot}$) in the Full Sample.}
    \label{fig:rave_spec_phot_comparison}
\end{figure*}

The rest of our sample is assembled from recent photometric estimates of temperature, luminosity classes, and metallicity for candidate stars from the RAVE DR6 survey based on the procedure described by \citet{Huang2021c}. This study made use of recalibrated zero-points \citep{Huang2021b} in the narrow- and medium-band photometry obtained by the Sky Mapper Southern Survey (SMSS; \citealt{Wolf2018}) Data Release 2 (DR2; \citealt{Onken2019}), along with broad-band photometry from Gaia EDR3 \citep{GaiaCollaboration2021}. The average errors adopted for each of the stellar parameters for the photometric portion of the sample are $\pm 62$ K for $T_{\text{eff}}$, and $\pm 0.13$\,dex for [Fe/H], with values for each star listed in Table~\ref{tab:initial_data_descript} in the Appendix (See \citealt{Huang2021c} for more information on the errors). \citet{Huang2021c} have compared the photometric estimates of $T_{\text{eff}}$ and [Fe/H] from their catalog with several medium-resolution studies (including stars from the Pristine Survey follow-up reported by \citealt{Aguado2019}, and from the Best \& Brightest sample reported by \citealt{Limberg2021c}) and high-resolution studies that employed near-IR spectroscopy from APOGEE DR14 \citep{Abolfathi2018} and DR16 \citep{Ahumada2020}, as well as optical high-resolution spectroscopy from a number of individual papers, and find generally excellent agreement. A total of $8205$ stars have available photometric estimates of effective temperature and metallicity within desired ranges ([Fe/H]$_{Phot}$ $\leq -0.8$ and $4250 \leq T_{\rm eff, Phot}$ (K) $\leq 7000$); $1303$ of these stars also have available medium-resolution spectra. The total number of stars available is $8675$, when duplicates are properly taken in consideration between the spectroscopic and photometric sources through the process discussed in the next section. 

We refer to these stars as the Full Sample. The spatial distribution of the Full Sample in Galactic coordinates can be seen in Figure~\ref{fig:galactic_map}. Figure~\ref{fig:mag_hist} compares the distributions of apparent magnitudes and colors for the Full Sample stars, as described in more detail below. Trimming of this sample to obtain a subset of stars suitable for our dynamical analysis is described below for the Initial Sample.

\subsection{Construction of the Initial Sample}

The metallicities from both RAVE DR6 and SMSS DR2 photometric estimates were available to construct the Initial Sample. Here we explore the comparison between the two metallicity sets and provide reasoning on our selection methods.

\subsubsection{Comparing RAVE DR6 and Photometric Metallicities}\label{subsubsec:metallicity_comp}

Comparisons of the metallicity estimates of the RAVE values to those with updated spectroscopic stellar parameters revealed a significant dispersion among the RAVE measurements, as can be appreciated in the left panel of Figure~\ref{fig:rave_spec_phot_comparison}. The comparison between the photometric estimates of the metallicity and the spectroscopic metallicity is shown in the same figure in the right panel. The biweight (see \citealt{Beers1990a}) estimates of offsets and scale in the residuals between the RAVE DR6 metallicities and the spectroscopic metallicities (in the sense [Fe/H]$_{RAVE}$ $-$ [Fe/H]$_{Spec}$) is $\mu = +0.32$ dex and $\sigma = 0.57$ dex, respectively, while the biweight offset and scale in the residuals between the photometric metallicities and spectroscopic metallicities (in the sense [Fe/H]$_{Phot}$ $-$ [Fe/H]$_{Spec}$) determined from the validation spectra is $\mu = +0.12$ dex and $\sigma = 0.30$ dex, respectively. Based on these behaviors, we decided to use the photometric determinations based on the recalibrated SMSS DR2 from \citet{Huang2021c} for the present analysis, rather than the RAVE DR6 estimates.

\subsubsection{Comparing Spectroscopic and Photometric Parameters}\label{subsubsec:spec_phot_comp}

\begin{figure*}
    \includegraphics[width=\textwidth]{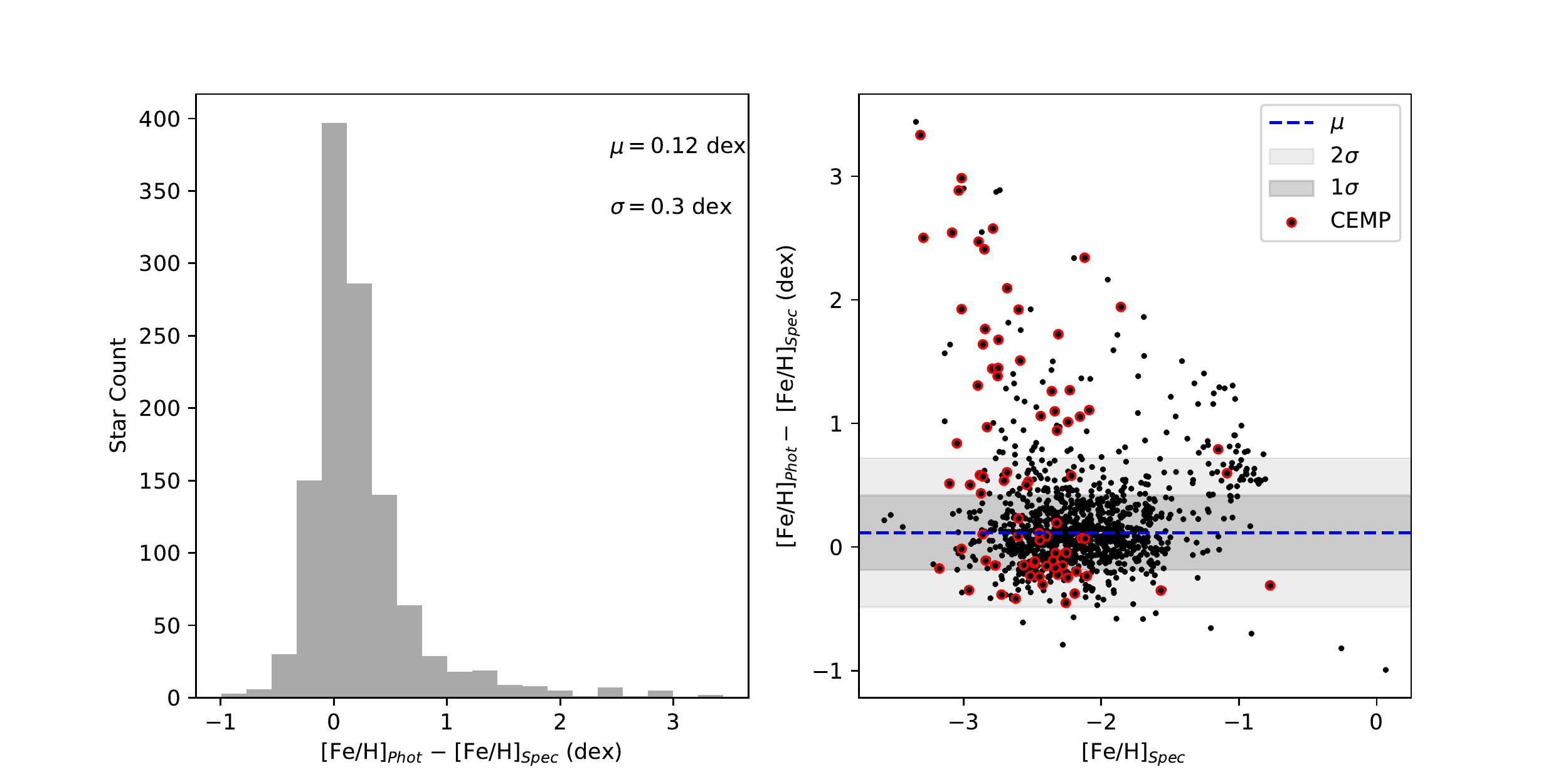}
    \caption{Left Panel: Histogram of the residuals of the difference between the photometric metallicity values ([Fe/H]$_{Phot}$) and the metallicity values obtained from medium-resolution spectra (Fe/H]$_{Spec}$) for the Full Sample. The biweight locations and scales are noted. Right panel: The residuals between [Fe/H]$_{Phot}$ and [Fe/H]$_{Spec}$, as a function of [Fe/H]$_{Spec}$, for the Full Sample. CEMP stars are indicated with red outlined points. The blue dashed line is the biweight location, while the shaded regions represent the first (1$\sigma$) and second (2$\sigma$) biweight scale ranges.  The apparent systematic discrepancy at high metallicity in this panel is likely due to difficulties in the n-SSPP estimates in this range (see text).}
    \label{fig:metallicity_diff_hist}
\end{figure*}

\begin{figure*}
    \includegraphics[width=\textwidth]{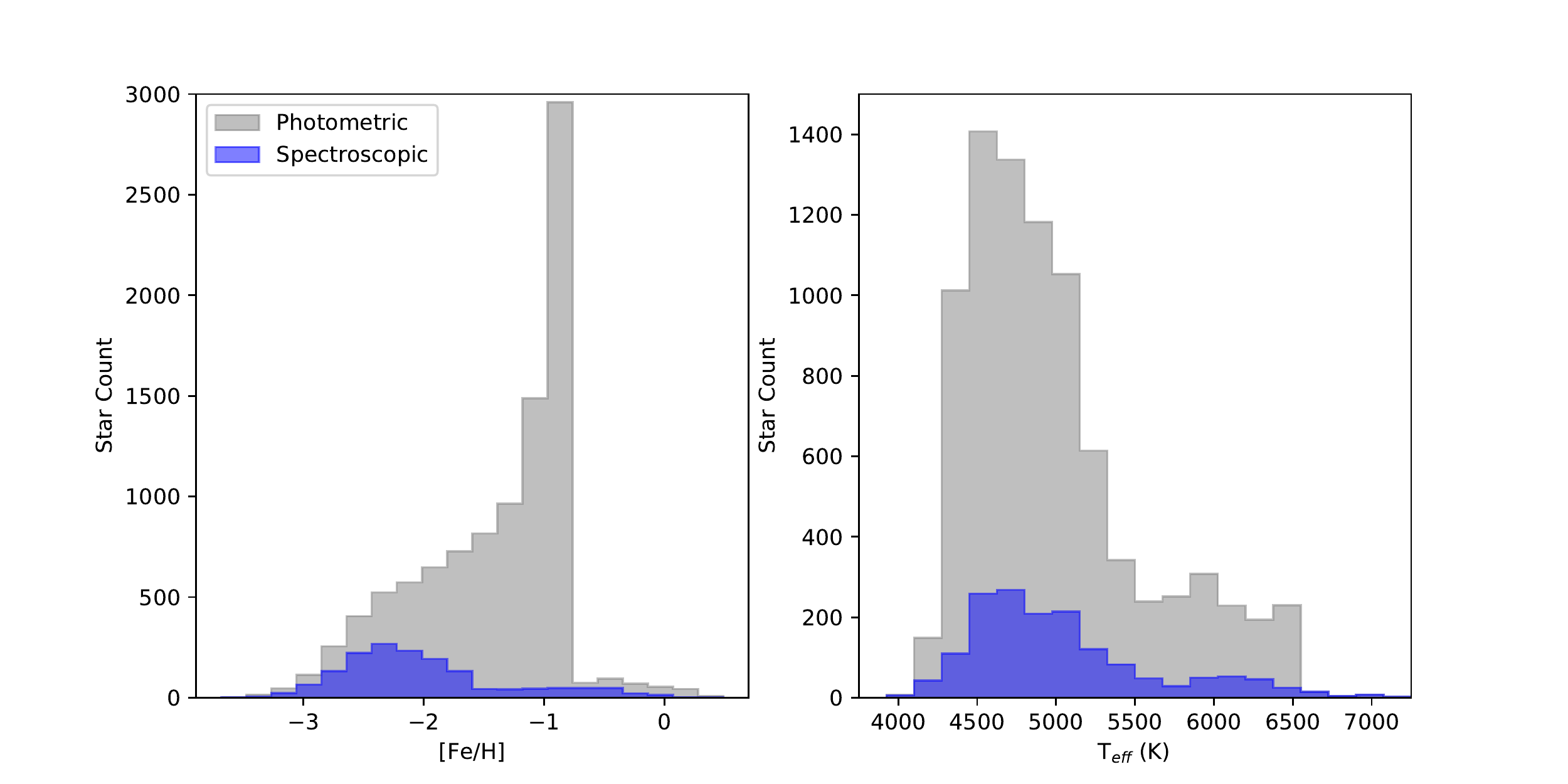}
    \caption{Left Panel: Histogram of the metallicity for the spectroscopic ([Fe/H]$_{Spec}$, blue) and photometric ([Fe/H]$_{Phot}$, gray) stars for the Full Sample. Right Panel: Histogram of the effective temperatures for the spectroscopic (T$_{\rm eff~Spec}$, blue) and photometric (T$_{\rm eff~Phot}$, gray) stars for the Full Sample.  Although the effective temperature distributions for the two subsets appear similar, the spectroscopic sample of stars clearly favors stars with lower [Fe/H], by design.}
    \label{fig:metallicity_teff_hist}
\end{figure*}

The comparison between metallicity determinations from both the spectroscopic and photometric estimates can be appreciated in Figure~\ref{fig:metallicity_diff_hist}. The biweight estimators of location and scale for the metallicity residuals determined from the medium-resolution spectra and the photometric metallicity yield $\mu =+0.12$ dex and $\sigma = 0.30$ dex. As noted by \citet{Huang2021c}, stars that have enhanced carbon often estimated photometric metallicities that are somewhat higher than the spectroscopic determinations, due to molecular carbon features affecting the blue narrow/medium-band filters $v$ and $u$ from SMSS (particularly for cooler carbon-enhanced stars).  Stars with [C/Fe]$_c > +0.7$, which we define as carbon-enhanced metal-poor (CEMP) stars are indicated with red circles around the dots shown in the bottom panel of Figure~\ref{fig:metallicity_diff_hist}. We note that, when these stars are removed from the sample, similar offsets and residuals are found as for the entire sample. 

\begin{figure}[t]
    \includegraphics[width=0.48\textwidth]{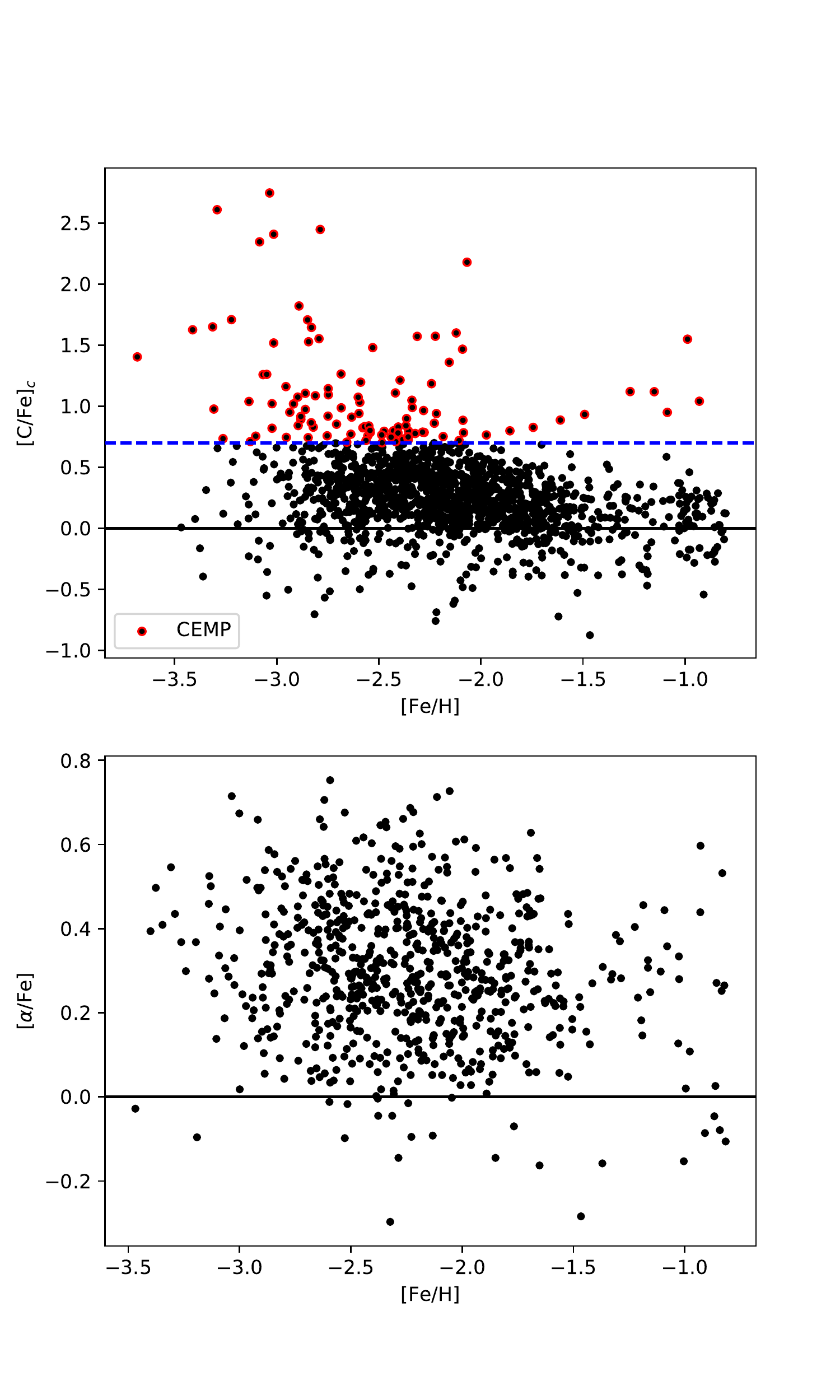}
    \caption{Top Panel: The corrected carbon abundance ([C/Fe]$_{c}$) of the Initial Sample, as a function of the final metallicity ([Fe/H]) for stars with [Fe/H] $\leq -0.8$. The CEMP cutoff ([C/Fe]$_{c} = +0.7$) is noted as the blue dashed line; CEMP stars are indicated with red outlined points. The Solar value is indicated as the solid black line. Bottom panel: The $\alpha$-element abundance ([$\alpha$/Fe]) of the Initial Sample, as a function of the final metallicity ([Fe/H]). The Solar value is indicated with a solid black line.}
    \label{fig:abundances}
\end{figure}

\begin{figure}[t]
    \includegraphics[width=0.48\textwidth]{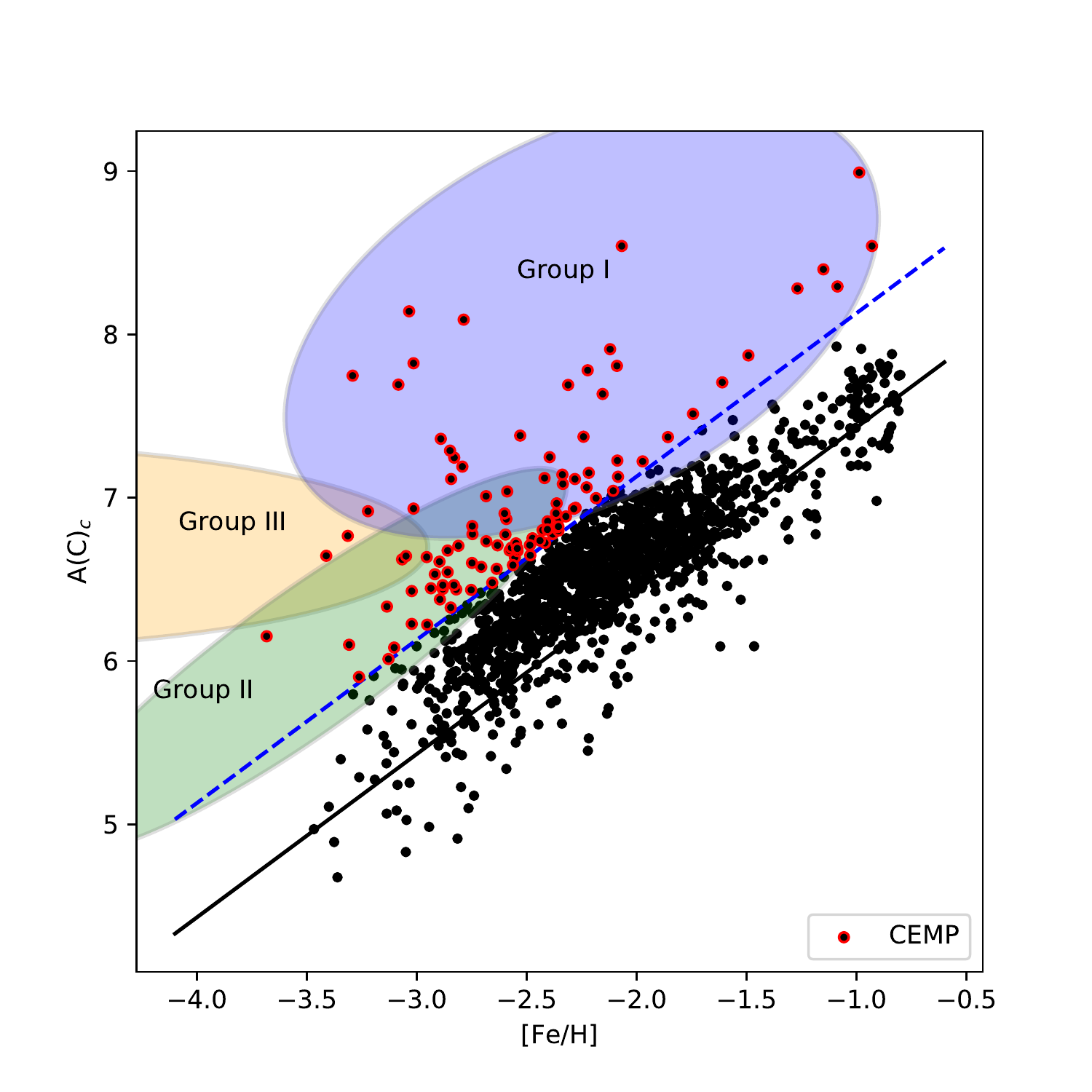}
    \caption{The Yoon-Beers Diagram of $A{\rm (C)}_c$, as a function of [Fe/H], for stars in the Initial Sample. The CEMP cutoff ([C/Fe]$_{c} = +0.7$) is indicated with a blue dashed line. The CEMP stars ([C/Fe]$_{c} > +0.7$) are shown as red outlined points. [C/Fe] $= 0$ is indicated with a solid black line. The ellipses represent the three different morphological groups of CEMP stars (See Figure~1 in \citealt{Yoon2016} for a comparison and more information).}
    \label{fig:yoon_beers}
\end{figure}

From inspection of this figure, there appears to be a systematic discrepancy in the metallicity estimates in the metal-rich region. As in Paper I, we attribute this to difficulties encountered by the n-SSPP estimates\footnote{The n-SSPP uses a subset of the estimators employed by the SSPP, which has been demonstrated previously to work very well in this high-metallicity range.}, rather than the photometric estimates, which have been shown by \citet{Huang2021c} to have excellent performance in this metallicity regime.  As noted below, we only retain stars with [Fe/H] $\leq -0.8$ in the Final Sample, so these stars will not greatly affect our subsequent analysis. 

Figure~\ref{fig:metallicity_teff_hist} shows the distributions of [Fe/H] and $T_{\rm eff}$ estimates obtained from the photometric and spectroscopic sub-samples.  As can be appreciated from inspection of this figure, although these subsets are similarly distributed over $T_{\rm eff}$ (right panel), the majority of the the stars in the spectroscopic validation sample have [Fe/H] $\lesssim -1.8$ (left panel), resulting from the selection of very low-metallicity candidates for the validation process.

\begin{figure}
    \includegraphics[width=0.5\textwidth]{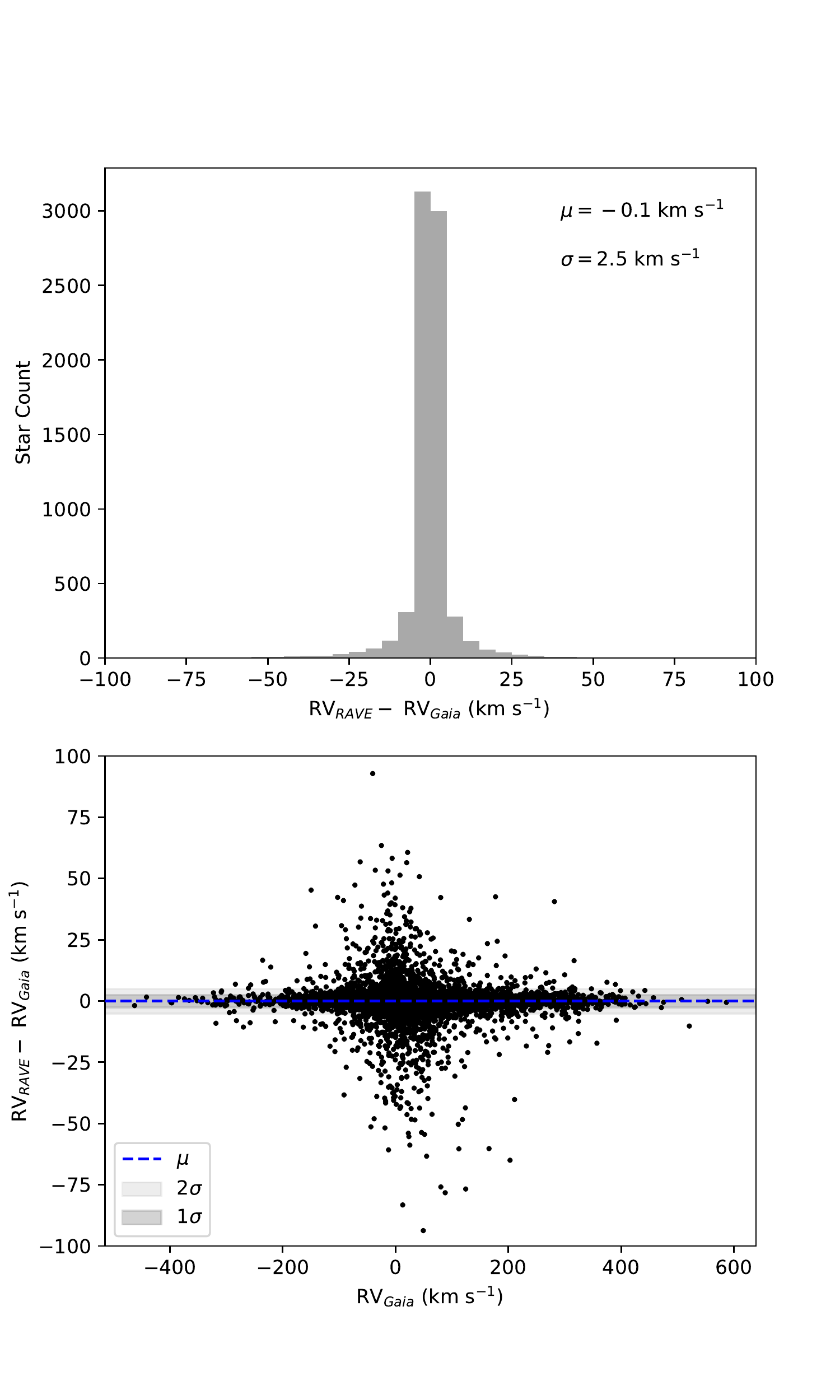}
    \caption{Top Panel: Histogram of the residuals of the difference between the RAVE DR6 radial velocities and the Gaia EDR3 values in the Full Sample. The biweight location and scale are noted. Bottom panel: The residuals between the RAVE DR6 and Gaia EDR3 radial velocities in the Full Sample, as a function of the Gaia radial velocities. The blue dashed line is the biweight location of the residual difference ($\mu = -0.1$ km s$^{-1}$), while the shaded regions represent the first (1$\sigma$ = 2.5 km s$^{-1}$), and second (2$\sigma$ = 5.0 km s$^{-1}$) biweight scale ranges.}
    \label{fig:rv_hist}
\end{figure}

\begin{figure}[t]
    \includegraphics[width=0.5\textwidth]{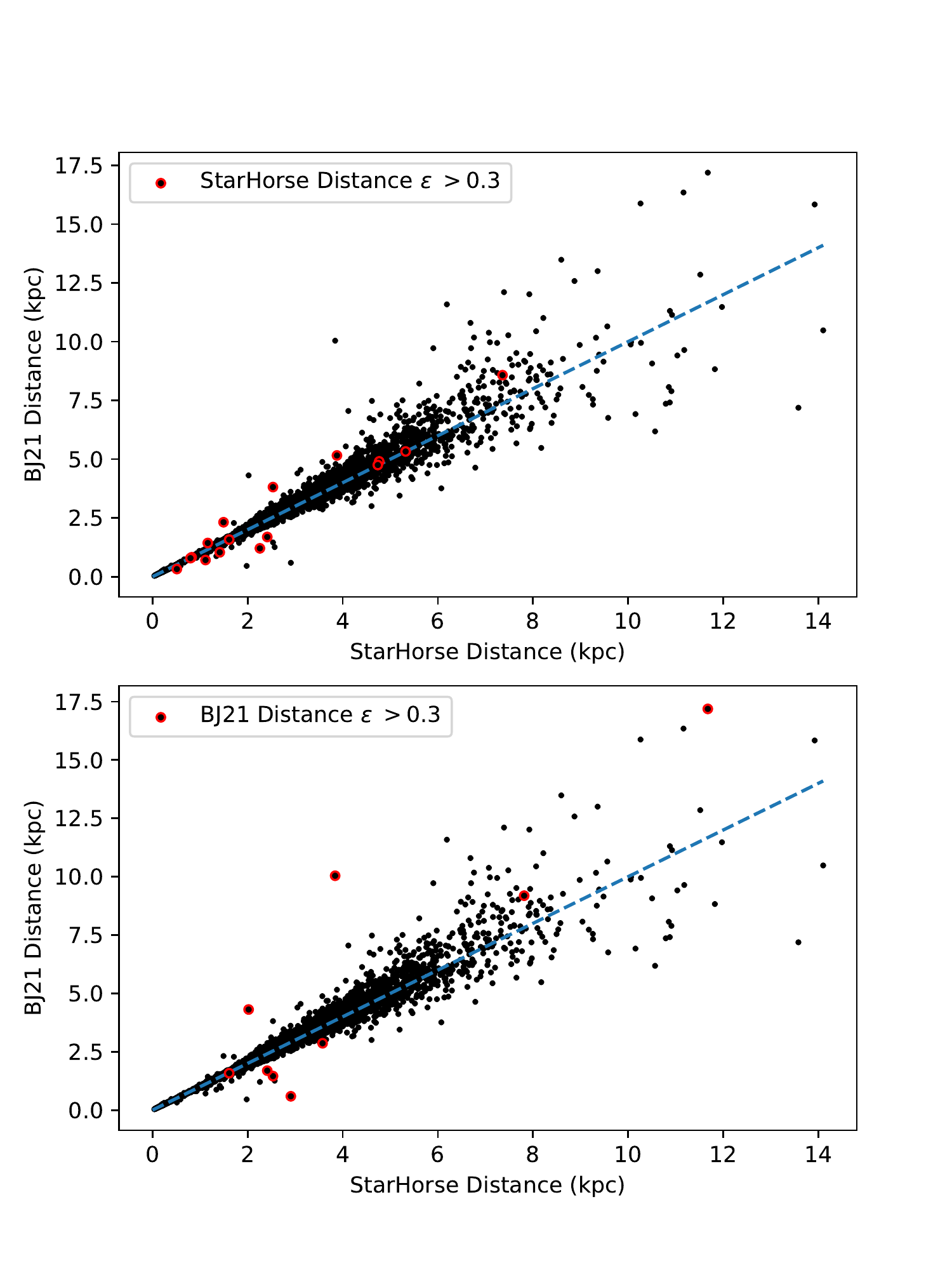}
    \caption{Top Panel: Comparison between the StarHorse distances and BJ21 distances in the Full Sample. Stars with the red outlined points indicate that the relative distance error in StarHorse is $\epsilon > 0.3$. Bottom panel: The same comparison as the top panel, but with the red outlined points indicating the relative distance error in BJ21 is $\epsilon > 0.3$. In both plots the dashed line indicates a one-to-one comparison between the two samples.}
    \label{fig:dist_comp}
\end{figure}

For stars that have both spectroscopic and photometric stellar parameters, we perform a procedure to obtain the parameters available to produce final adopted estimates of [Fe/H] and $T_{\rm eff}$, as in Paper I. If the absolute difference in metallicity between the two samples ([Fe/H]$_{Spec}$ $-$ [Fe/H]$_{Phot}$) is less than $0.5$ dex, then we average the two sets of parameters. If not, then a choice is made to adopt either the spectroscopic or photometric parameters based on visual inspection of the spectrum for a given star. For the stars with available spectroscopy, we then adjust estimates of the carbon and $\alpha$-element abundances from the n-SSPP based on the final adopted [Fe/H].

Restricting stellar values to [Fe/H] $\leq -0.8$ and $4250 \leq T_{\rm eff} \,\rm{(K)} \leq 7000$ (consistent with the choices made in Paper I) yields $8377$ metal-poor stars for the Initial Sample.

The [C/Fe]$_{c}$ and [$\alpha$/Fe] estimates for stars in the Initial Sample are shown in Figure~\ref{fig:abundances}. The Yoon-Beers diagram of A$(C)_{c}$ vs. [Fe/H] for Initial Sample stars is shown in Figure~\ref{fig:yoon_beers}.  The CEMP stars ([C/Fe]$_c > +0.7$) are provided in Table~\ref{tab:cemp} of the Appendix.

The stars from the Full Sample were then cross-matched with Gaia Early Data Release 3 (EDR3; \citealt{GaiaCollaboration2021}) using a $5 \arcsec$ radius to find their $6$-D astrometric parameters. To validate the match for each star, confirmation was performed by checking that the stellar magnitudes agreed to within $0.5$ mag between the sources. The Full Sample was mostly taken from \textit{V} magnitudes supplied by the AAVSO Photometric All Sky Survey (APASS; \citealt{Henden2014}) Data Release 9 (DR9; \citealt{Henden2016}), with various other sources listed in the Appendix tables supplying the rest. The corresponding matches were then compared with the \textit{V} magnitude utilizing the transformation from Gaia magnitudes \textit{G}, \textit{G}$_{\text{BP}}$, and \textit{G}$_{\text{RP}}$ in EDR3 provided in Table C.2 of \citet{Riello2021}. A comparison of the magnitudes and colors for the photometric and spectroscopic subsets can be seen in Figure \ref{fig:mag_hist}. From inspection of the figure, although the relative numbers differ, the distributions are similar to one another. 

\subsection{Construction of the Final Sample}\label{subsec:final_sample}

Once stars from  the Full Sample are matched with those from Gaia EDR3, dynamical parameters for stars in the Initial Sample can be recovered and used to construct the Final Sample.

\subsubsection{Radial Velocities, Distances, and Proper Motions}

Radial velocities, parallaxes, and proper motions for each star are taken from Gaia EDR3, when available. Note that the radial velocities for the stars in Gaia EDR3 are available for about $87\%$ of the Full Sample. Typical errors for Gaia EDR3 radial velocities are ${\sim} 1$ km s$^{-1}$. The top panel of Figure~\ref{fig:rv_hist} shows a histogram of the residual differences between the RAVE DR6 radial velocities and the Gaia EDR3 values. The biweight location and scale of these differences are $\mu = -0.1$ km s$^{-1}$ and $\sigma = 2.5$ km s$^{-1}$, respectively. The bottom panel of this figure shows the residuals between the RAVE DR6 and Gaia EDR3 radial velocities, as a function of the Gaia radial velocities. The blue dashed line is the biweight location, while the shaded regions represent the first (1$\sigma$) and second (2$\sigma$) biweight scale ranges.  We expect that many of the stars with residuals outside the 2$\sigma$ range are binaries, causing an improper measurement of the systemic radial velocities.

\begin{figure}
    \includegraphics[width=0.49\textwidth]{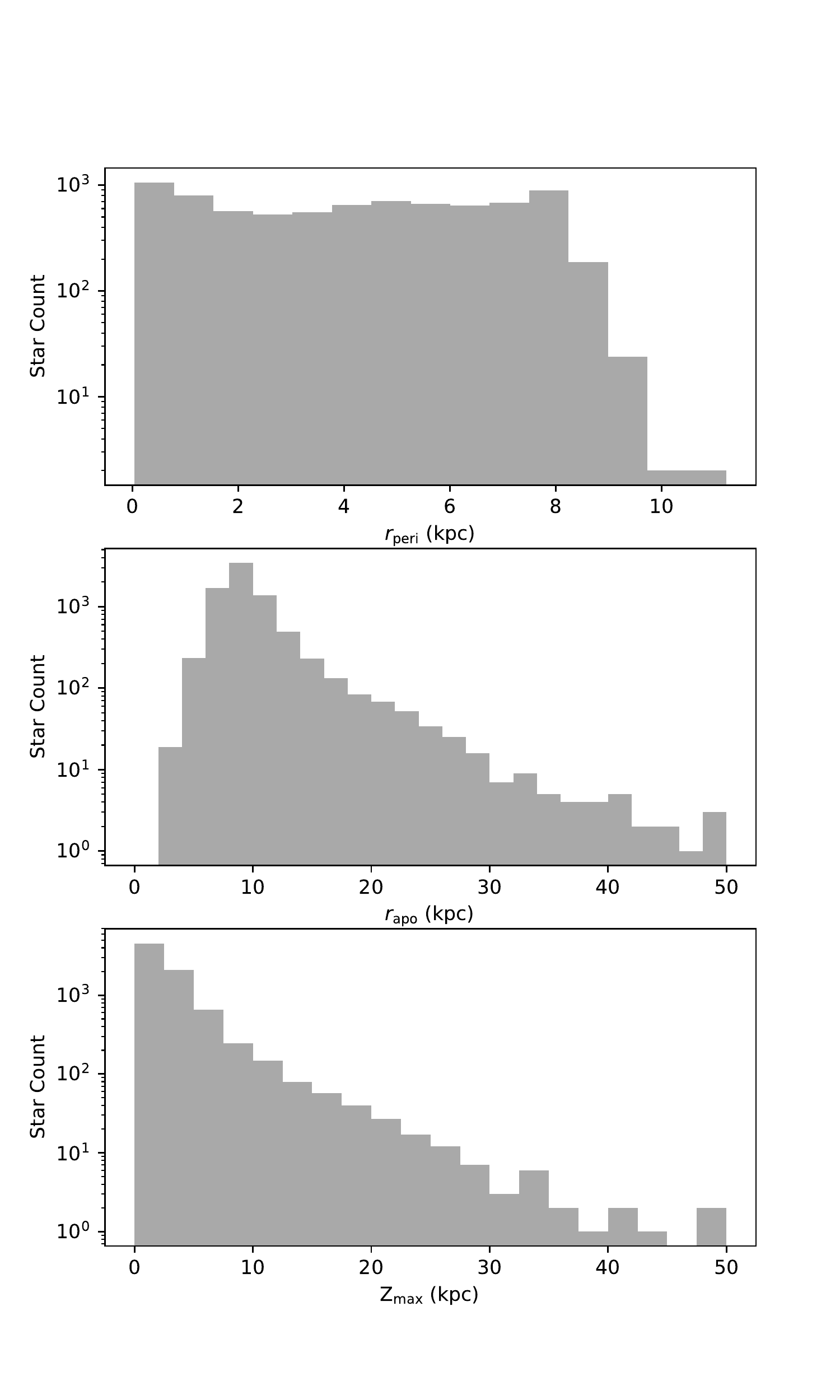}
    \caption{Logarithmic histograms of the orbital parameters \textit{r}$_{\peri}$ (top), \textit{r}$_{\apo}$ (middle), and Z$_{\maxtext}$ (bottom) for the RAVE DR6 Final Sample. Note that a few stars have \textit{r}$_{\apo}$ and Z$_{\maxtext}$ outside the range shown in the panels.}
    \label{fig:orb_dist}
\end{figure}

The distances to the stars are determined either through the StarHorse distance estimate \citep{Anders2021} or the Bailer-Jones distance estimate (BJ21; \citealt{Bailer-Jones2021}). Parallax values in our Full Sample from EDR3 have an average error of around $0.04$ mas. The StarHorse and BJ21 distances are determined by a Bayesian approach utilizing the EDR3 parallax, magintude, and color \citep{Anders2021,Bailer-Jones2021}. The errors are presented for each star in the tables provided in the Appendix. We prioritize the StarHorse distances when the relative error (the error divided by the reported value), $\epsilon$, is $ \epsilon < 30\%$. If the StarHorse relative error is $\geq 30\%$, then we adopt the BJ21 distance if the relative error is $ \epsilon < 30\%$. If only one distance estimator is available, then we adopt it. If both distances, or the only available distance, have $\epsilon \geq 30\%$, then we discard the star from the dynamical analysis below. Note that in Figure~\ref{fig:dist_comp} the StarHorse and BJ21 approaches produce similar distances, especially when the distance is smaller than $5$ kpc. A comparison between Figure~\ref{fig:dist_comp} and Figure 8 in Paper I clearly shows the improvement in the distance estimators when the StarHorse distances are used in comparison to the Gaia EDR3 inverse-parallax distances employed in Paper I. The proper motions in our Full Sample from Gaia EDR3 have an average error of $39$ $\mu$as yr$^{-1}$.

\subsection{Dynamical Parameters}\label{subsec:DynamicalParameters}

The orbital characteristics of the stars are determined using the Action-based GAlaxy Modelling Architecture\footnote{\url{http://github.com/GalacticDynamics-Oxford/Agama}} (\AGAMA) package \citep{Vasiliev2019}, using the same Solar positions and peculiar motions described in Paper I\footnote{We adopt a Solar position of ($-8.249$, $0.0$, $0.0$) kpc \citep{GravityCollaboration2020} and Solar peculiar motion in (U,W) as ($11.1$,$7.25$) km s$^{-1}$ \citep{Schonrich2010} with Galoctocentric Solar azimuthal velocity  \textit{V} $= 250.70$ km s$^{-1}$ determined from \citet{Reid2020}.}. The MW gravitational potential we adopt is the \MWMMXVII ~potential \citep{McMillan2017}, also described in Paper I. The 6$D$ astrometric parameters, determined in Section~\ref{sec:Data}, are run through the orbital integration process in \AGAMA ~using the same procedure outlined in Paper I to calculate the orbital energy, cylindrical positions and velocities, angular momentum, cylindrical actions, and eccentricity. See Paper I for definitions of these orbital parameters.

The above procedure obtains the orbital parameters if the astrometric parameters are precisely described by the given values. However, these values have errors associated with them, so a method must be developed to estimate the errors in the orbital parameters. This is accomplished through a Monte Carlo sampling over the errors in the astrometric parameters. The procedure that we employ to determine the orbital errors using Monte Carlo sampling is described in detail in Paper I. 

An inspection is performed to identify stars that are not suitable for the following dynamical analysis. The Initial Sample of $8377$ stars contains stars that are unbound from the MW (E$ > 0$), along with stars that do not have the full $6$-D astrometric parameters of position, radial velocity, distance, and proper motions were then cut. Finally, in order to obtain accurate orbital dynamics, we remove $401$ stars with differences in their RAVE DR6 radial velocities compared to the Gaia radial velocities that lie outside the $15$ km s$^{-1}$ range. Most of these stars are expected to be binaries. Application of this cut leaves a total sample of $7957$ stars to perform the following analysis. The dynamical parameters of the stars with orbits determined are listed in Table~\ref{tab:final_data_descript} in the Appendix; we refer to this as the Final Sample.

\begin{deluxetable*}{l  r  r  c}
\tabletypesize{\scriptsize}
\tablecaption{Identified DTGs \label{tab:cluster_summary}}
\tablehead{\colhead{DTG} & \colhead{$N$ Stars} & \colhead{Confidence} & \colhead{Associations}}
\startdata
1 & $35$ & $99.0\%$ & new\\
2 & $30$ & $79.0\%$ & DS21:DTG-13\\
3 & $30$ & $93.0\%$ & new\\
4 & $28$ & $90.5\%$ & DS21:DTG-4, DG21:CDTG-18, GM18a:C1, HL19:GL-2, GL21:DTG-8\\
5 & $27$ & $77.1\%$ & Splashed Disk\\
6 & $24$ & $59.1\%$ & DS21:DTG-47\\
7 & $23$ & $87.7\%$ & GSE, IR18:E, DG21:CDTG-1, EV21:NGC~4833, GC21:Sausage, GL21:DTG-37, DS21:DTG-43\\
8 & $23$ & $95.5\%$ & GSE, DS21:DTG-3, GC21:Sausage, DG21:CDTG-9, GL21:DTG-34\\
9 & $23$ & $86.4\%$ & new\\
10 & $22$ & $98.9\%$ & DS21:DTG-12\\
\enddata
\tablecomments{We adopt the nomenclature for previously identified DTGs and CDTGs from \cite{Yuan2020b}.}\tablecomments{This table is a stub; the full table is available in the electronic edition.}\end{deluxetable*}

Figure~\ref{fig:orb_dist} provides histograms of \textit{r}$_{\apo}$ (top), \textit{r}$_{\peri}$ (middle), and Z$_{\maxtext}$ (bottom) for the Final Sample.  From inspection of this figure, it is clear that the majority of the stars in this sample occupy orbits that take them inside the inner-halo region, but they also explore regions well into the outer-halo region, up to $\sim 50$ kpc away.

\section{Clustering Procedure}\label{sec:ClusteringProcedure}

\citet{Helmi2000} were among the first to suggest the use of integrals of motion, in their case orbital energy and angular momenta, to find substructure in the MW using the precision measurements of next-generation surveys that were planned at the time. \citet{McMillan2008} suggested the use of actions as a complement to the previously suggested orbital energy and angular momenta, with only the vertical angular momentum being invariant in an axisymmetric potential. Most recently, many authors have employed the orbital energy and cylindrical actions (E,J$_{r}$,J$_{\phi}$,J$_{z}$) to determine the substructure of the MW using Gaia measurements \citep{Helmi2017,Myeong2018b,Myeong2018c,Roederer2018a,Naidu2020,Yuan2020a,Yuan2020b,Gudin2021,Limberg2021a,Shank2021}.

As described in Paper I, we employ \HDBSCAN ~in order to perform a cluster analysis over the orbital energy and cylindrical actions from the Final Sample obtained through the procedure outlined in Section \ref{subsec:final_sample}. The \HDBSCAN ~algorithm\footnote{For a detailed description of the \HDBSCAN ~algorithm visit: \url{https://hdbscan.readthedocs.io/en/latest/how_hdbscan_works.html}} operates through a series of calculations that are able to separate the background noise from denser clumps of data in the dynamical parameters. We utilize the following parameters described in Paper I: \verb ~min_cluster_size ~$= 5$, \verb ~min_samples ~$= 5$, \verb ~cluster_selection_method ~$=$ \verb 'leaf' , \verb ~prediction_data ~$=$ \verb ~True ~, Monte Carlo samples at $1000$, and minimum confidence set to $20\%$.

Table~\ref{tab:cluster_summary} provides a listing of the Dynamically Tagged Groups (DTGs) identified by this procedure, along with their numbers of member stars, confidence values, and associations described below.  Note that, although a minimum confidence value of $20\%$ was employed, the actual minimum value found for these DTGs is $23.9\%$. The DTGs and CDTGs are identified using the nomenclature introduced by \citet{Yuan2020b}, to which we refer the interested reader. 

Table~\ref{tab:cluster_results_stub} lists the stellar members of the identified DTGs, along with their final values of [Fe/H], [C/Fe], [C/Fe]$_c$, and [$\alpha$/Fe], where available.  The last line in the listing for each DTG gives the mean and dispersion (both using biweight estimates) for each quantity.

Table~\ref{tab:cluster_results_element_stub} lists the identified DTGs where at least half of the member stars have [C/Fe] and [$\alpha$/Fe] detected. 

Table~\ref{tab:cluster_orbit_stub} lists the derived dynamical parameters derived by \AGAMA ~used in our analysis.

\section{Structure Associations}\label{sec:StructureAssociations}

Associations between the newly identified DTGs are now sought between known MW structures, including large-scale substructures, previously identified dynamical groups, stellar associations, globular clusters, and dwarf galaxies.

\begin{deluxetable*}{l  c  c  c  c}
\tablecaption{DTGs Identified by \HDBSCAN \label{tab:cluster_results_stub}}
\tablehead{\colhead{RAVE ID} & \colhead{[Fe/H]} & \colhead{[C/Fe]} & \colhead{[C/Fe]$_{c}$} & \colhead{[$\alpha$/Fe]}}
\startdata
\multicolumn{5}{c}{$DTG-1$} \\
\multicolumn{5}{c}{Structure: Unassigned Structure} \\
\multicolumn{5}{c}{Group Assoc: No Group Associations} \\
\multicolumn{5}{c}{Stellar Assoc: No Stellar Associations} \\
\multicolumn{5}{c}{Globular Assoc: No Globular Associations} \\
\multicolumn{5}{c}{Dwarf Galaxy Assoc: No Dwarf Galaxy Associations} \\
J124828.60$-$463558.7 & $-0.910$ & $\dots$ & $\dots$ & $\dots$ \\
J134259.33$-$420116.0 & $-0.970$ & $\dots$ & $\dots$ & $\dots$ \\
J145237.82$-$394929.1 & $-0.860$ & $\dots$ & $\dots$ & $\dots$ \\
J145301.90$-$734037.1 & $-1.390$ & $\dots$ & $\dots$ & $\dots$ \\
J145314.34$-$255954.2 & $-0.860$ & $\dots$ & $\dots$ & $\dots$ \\
J145848.84$-$380905.7 & $-1.000$ & $\dots$ & $\dots$ & $\dots$ \\
J153735.89$-$025256.1 & $-0.840$ & $\dots$ & $\dots$ & $\dots$ \\
J160708.39$-$025922.7 & $-0.970$ & $\dots$ & $\dots$ & $\dots$ \\
J161555.25$-$004906.2 & $-1.200$ & $\dots$ & $\dots$ & $\dots$ \\
J161821.47$-$023635.3 & $-0.820$ & $\dots$ & $\dots$ & $\dots$ \\
J162133.93$-$075800.0 & $-1.090$ & $\dots$ & $\dots$ & $\dots$ \\
J162608.99$-$062248.1 & $-1.080$ & $\dots$ & $\dots$ & $\dots$ \\
J182742.83$-$513704.2 & $-2.170$ & $\dots$ & $\dots$ & $\dots$ \\
J184219.23$-$754942.7 & $-0.920$ & $\dots$ & $\dots$ & $\dots$ \\
J190528.57$-$384355.5 & $-2.310$ & $\dots$ & $\dots$ & $\dots$ \\
J191626.47$-$630945.4 & $-0.900$ & $\dots$ & $\dots$ & $\dots$ \\
J191718.05$-$422357.3 & $-0.880$ & $\dots$ & $\dots$ & $\dots$ \\
J191738.19$-$602128.7 & $-0.900$ & $\dots$ & $\dots$ & $\dots$ \\
J195053.84$-$401254.9 & $-1.570$ & $\dots$ & $\dots$ & $\dots$ \\
J195603.92$-$534035.6 & $-0.800$ & $\dots$ & $\dots$ & $\dots$ \\
J195610.10$-$452825.4 & $-1.080$ & $\dots$ & $\dots$ & $\dots$ \\
J195924.29$-$282505.1 & $-0.920$ & $\dots$ & $\dots$ & $\dots$ \\
J200012.26$-$582218.7 & $-0.940$ & $\dots$ & $\dots$ & $\dots$ \\
J200344.00$-$274647.1 & $-0.870$ & $\dots$ & $\dots$ & $\dots$ \\
J202109.85$-$153156.4 & $-0.920$ & $\dots$ & $\dots$ & $\dots$ \\
J202128.83$-$392854.3 & $-1.180$ & $\dots$ & $\dots$ & $\dots$ \\
J203015.48$-$353149.9 & $-0.900$ & $\dots$ & $\dots$ & $\dots$ \\
J203540.97$-$082246.7 & $-0.950$ & $\dots$ & $\dots$ & $\dots$ \\
J204434.37$-$144211.0 & $-0.810$ & $\dots$ & $\dots$ & $\dots$ \\
J205829.82$-$181249.9 & $-1.660$ & $\dots$ & $\dots$ & $\dots$ \\
J210531.08$-$025632.7 & $-1.280$ & $\dots$ & $\dots$ & $\dots$ \\
J210711.66$-$160721.0 & $-0.840$ & $\dots$ & $\dots$ & $\dots$ \\
J213042.25$-$151010.3 & $-0.810$ & $\dots$ & $\dots$ & $\dots$ \\
J215617.20$-$180737.8 & $-0.890$ & $\dots$ & $\dots$ & $\dots$ \\
J220930.99$-$261800.8 & $-0.880$ & $\dots$ & $\dots$ & $\dots$ \\
$\mu \pm \sigma$ ([X/Y]) & $-0.921\pm0.132$ & $\dots$ & $\dots$ & $\dots$\\
\pagebreak
\enddata
\tablecomments{$\mu$ and $\sigma$ represent the biweight estimates of the location and scale for the abundances in the DTG.}\tablecomments{This table is a stub; the full table is available in the electronic edition.}\end{deluxetable*}

\subsection{Milky Way Substructures}\label{subsec:MWSubstructure}

\begin{deluxetable*}{l  c  c  c  c}
\tablecaption{DTGs Identified by \HDBSCAN ~with Carbon and $\alpha$-Elements \label{tab:cluster_results_element_stub}}
\tablehead{\colhead{} & \colhead{[Fe/H]} & \colhead{[C/Fe]} & \colhead{[C/Fe]$_{c}$} & \colhead{[$\alpha$/Fe]}}
\startdata
\multicolumn{5}{c}{$DTG-4$ [N$_{C}$: (14/28) and N$_{\alpha}$: (11/28)]} \\
\multicolumn{5}{c}{Structure: Unassigned Structure} \\
\multicolumn{5}{c}{Group Assoc: C1: \citet{Myeong2018b}} \\
\multicolumn{5}{c}{Group Assoc: GL-2: \citet{Li2019}} \\
\multicolumn{5}{c}{Group Assoc: DTG-8: \citet{Limberg2021a}} \\
\multicolumn{5}{c}{Group Assoc: DTG-4: \citet{Shank2021}} \\
\multicolumn{5}{c}{Stellar Assoc: J113738.01-520224.8 (DTG-4: \citet{Shank2021})} \\
\multicolumn{5}{c}{Stellar Assoc: 2MASS J15075699-0659593 (CDTG-18: \citet{Gudin2021})} \\
\multicolumn{5}{c}{Stellar Assoc: J225118.77-381438.6 (DTG-4: \citet{Shank2021})} \\
\multicolumn{5}{c}{Stellar Assoc: 2MASS J23362202-5607498* (CDTG-18: \citet{Gudin2021})} \\
\multicolumn{5}{c}{Globular Assoc: No Globular Associations} \\
\multicolumn{5}{c}{Dwarf Galaxy Assoc: No Dwarf Galaxy Associations} \\
$\mu \pm \sigma$ ([X/Y]) & $-2.074\pm0.595$ & $-0.284\pm0.451$ & $+0.357\pm0.286$ & $+0.351\pm0.157$\\
\\
\multicolumn{5}{c}{$DTG-47$ [N$_{C}$: (5/11) and N$_{\alpha}$: (5/11)]} \\
\multicolumn{5}{c}{Structure: Thamnos} \\
\multicolumn{5}{c}{Group Assoc: Sequoia: \citet{Cordoni2021}} \\
\multicolumn{5}{c}{Stellar Assoc: TYC 8420-259-1 (VelHel-4: \citet{Helmi2017})} \\
\multicolumn{5}{c}{Globular Assoc: No Globular Associations} \\
\multicolumn{5}{c}{Dwarf Galaxy Assoc: No Dwarf Galaxy Associations} \\
$\mu \pm \sigma ([X/Y])$ & $-1.829\pm0.592$ & $-0.038\pm0.567$ & $+0.269\pm0.165$ & $+0.434\pm0.224$\\
\\
\multicolumn{5}{c}{$DTG-67$ [N$_{C}$: (8/10) and N$_{\alpha}$: (8/10)]} \\
\multicolumn{5}{c}{Structure: LMS-1 (Wukong)} \\
\multicolumn{5}{c}{Group Assoc: Comoving: \citet{Myeong2017}} \\
\multicolumn{5}{c}{Group Assoc: C1: \citet{Myeong2018b}} \\
\multicolumn{5}{c}{Group Assoc: Cand11: \citet{Myeong2018c}} \\
\multicolumn{5}{c}{Group Assoc: CDTG-4: \citet{Gudin2021}} \\
\multicolumn{5}{c}{Group Assoc: DTG-2: \citet{Limberg2021a}} \\
\multicolumn{5}{c}{Stellar Assoc: 2MASS J00182832-3900338 (CDTG-4: \citet{Gudin2021})} \\
\multicolumn{5}{c}{Stellar Assoc: 2MASS J00413026-4058547 (CDTG-4: \citet{Gudin2021})} \\
\multicolumn{5}{c}{Stellar Assoc: 2MASS J00453930-7457294 (CDTG-4: \citet{Gudin2021})} \\
\multicolumn{5}{c}{Globular Assoc: No Globular Associations} \\
\multicolumn{5}{c}{Dwarf Galaxy Assoc: No Dwarf Galaxy Associations} \\
$\mu \pm \sigma$ ([X/Y]) & $-2.386\pm0.404$ & $-0.177\pm0.537$ & $+0.263\pm0.317$ & $+0.379\pm0.170$\\
\\
\multicolumn{5}{c}{$DTG-71$ [N$_{C}$: (5/9) and N$_{\alpha}$: (4/9)]} \\
\multicolumn{5}{c}{Structure: Gaia-Sausage-Enceladus (GSE)} \\
\multicolumn{5}{c}{Group Assoc: DTG-37: \citet{Limberg2021a}} \\
\multicolumn{5}{c}{Stellar Assoc: TYC 5830-840-1 (VelHel-6: \citet{Helmi2017})} \\
\multicolumn{5}{c}{Globular Assoc: No Globular Associations} \\
\multicolumn{5}{c}{Dwarf Galaxy Assoc: No Dwarf Galaxy Associations} \\
$\mu \pm \sigma$ ([X/Y]) & $-1.780\pm0.362$ & $-0.074\pm0.237$ & $+0.253\pm0.005$ & $+0.320\pm0.140$\\
\\
\multicolumn{5}{c}{$DTG-78$ [N$_{C}$: (6/9) and N$_{\alpha}$: (4/9)]} \\
\multicolumn{5}{c}{Structure: Gaia-Sausage-Enceladus (GSE)} \\
\multicolumn{5}{c}{Group Assoc: CDTG-14: \citet{Gudin2021}} \\
\multicolumn{5}{c}{Stellar Assoc: 2MASS J00541965-0611555* (CDTG-14: \citet{Gudin2021})} \\
\multicolumn{5}{c}{Stellar Assoc: 2MASS J08155667-2204105 (CDTG-14: \citet{Gudin2021})} \\
\multicolumn{5}{c}{Stellar Assoc: 2MASS J10044858-2706500* (CDTG-14: \citet{Gudin2021})} \\
\multicolumn{5}{c}{Globular Assoc: No Globular Associations} \\
\multicolumn{5}{c}{Dwarf Galaxy Assoc: No Dwarf Galaxy Associations} \\
$\mu \pm \sigma$ ([X/Y]) & $-2.222\pm0.349$ & $+0.004\pm0.365$ & $+0.519\pm0.154$ & $+0.392\pm0.221$\\
\\
\enddata
\tablecomments{This table lists the DTGs where around half of the member stars have both [C/Fe] and [$\alpha$/Fe] detected. N$_{C}$ (N$_{\alpha}$) indicates the total number of Carbon ($\alpha$) detections in comparison to the total number of stars in the DTG.}\tablecomments{$\mu$ and $\sigma$ represent the biweight estimates of the location and scale for the abundances in the DTG.}\end{deluxetable*}

\begin{figure*}[t]
    \centering
    \includegraphics[width=0.98\textwidth,height=0.98\textheight,keepaspectratio]{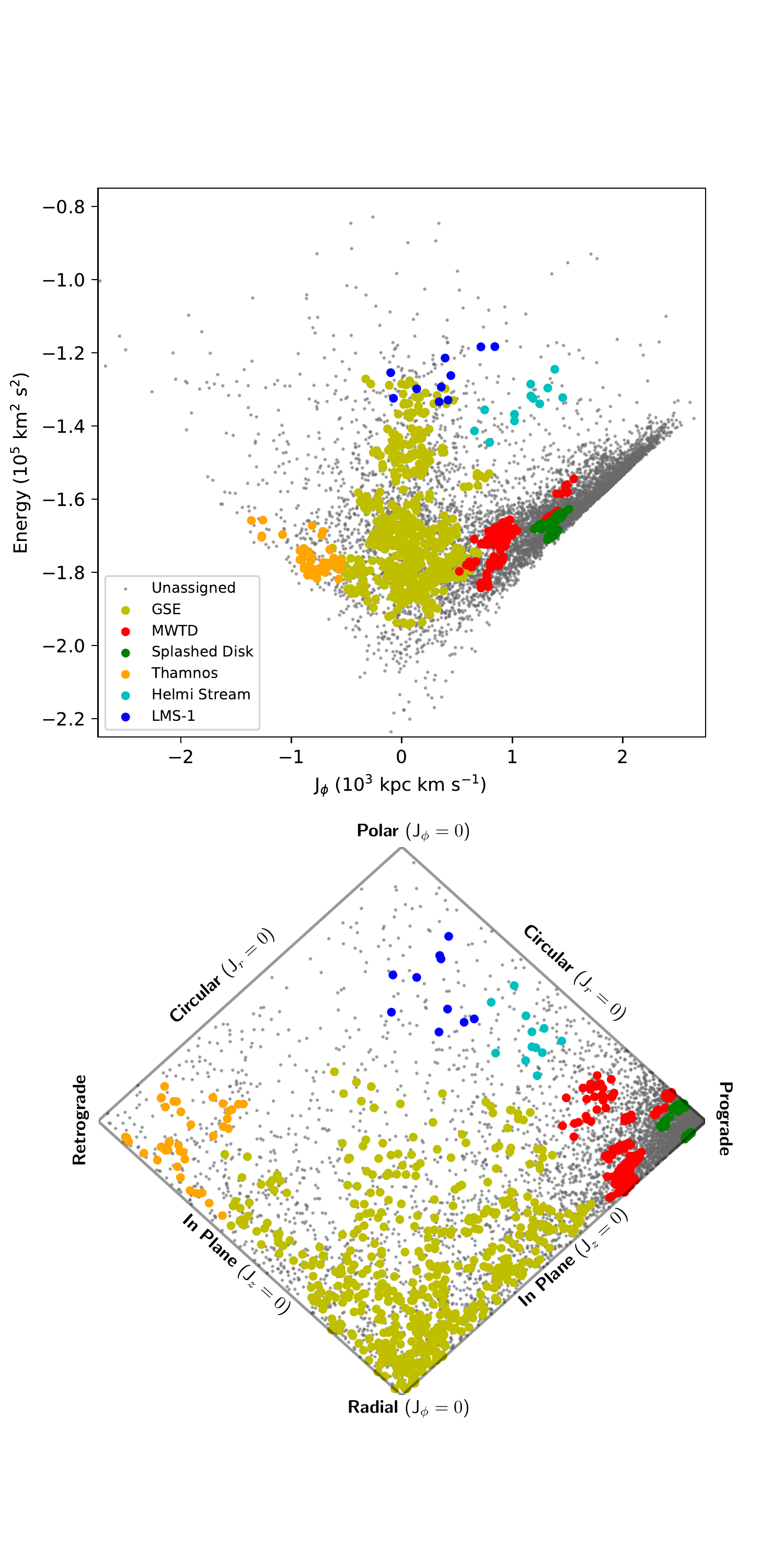}
    \caption{Top Panel: Lindblad Diagram of the identified MW substructures. The different structures are associated with the colors outlined in the legend. Bottom Panel: The projected-action plot of the same substructures. This space is represented by J$_{\phi}$/J$_{\text{Tot}}$ for the horizontal axis and (J$_{\text{z}}$ - J$_{\text{r}}$)/J$_{\text{Tot}}$ for the vertical axis with J$_{\text{Tot}}$ = J$_{\text{r}}$ + $|$J$_{\phi}|$ + J$_{\text{z}}$. For more details on the projected-action space, see Figure~3.25 in \citet{Binney2008}.}
    \label{fig:energy_actions}
\end{figure*}

Analyzing the orbital energy and actions is not sufficient to determine separate large-scale substructures. Information on the elemental abundances is crucial due to the differing star-formation histories of the structures, which can vary in both mass and formation redshift \citep{Naidu2020}. The outline for the prescription used to determine the structural associations with our DTGs is described in \citet{Naidu2020}, and explained in detail in Paper I. Simple selections are performed based on physically motivated choices for each substructure, excluding previously defined substructures, as the process iterates to decrease contamination between substructures. Following their procedures, we find $6$ predominant MW substructures associated with our DTGs, listed in Table~\ref{tab:substructures}. This table provides the numbers of stars in each substructure, the mean and dispersion of their chemical abundances, and the mean and dispersion of their dynamical parameters for each substructure.  The Lindblad diagram and projected-action plot for these substructures is shown in Figure~\ref{fig:energy_actions}.

\begin{deluxetable*}{l  r  c  c  c  r}
\tablecaption{Cluster Dynamical Parameters Determined by \AGAMA \label{tab:cluster_orbit_stub}}
\tablehead{\colhead{Cluster} & \colhead{$N$ Stars} & \colhead{($\langle$v$_{\text{r}}\rangle$,$\langle$v$_{\phi}\rangle$,$\langle$v$_{\text{z}}\rangle$)} & \colhead{($\langle$J$_{\text{r}}\rangle$,$\langle$J$_{\phi}\rangle$,$\langle$J$_{\text{z}}\rangle$)} & \colhead{$\langle$E$\rangle$} & \colhead{$\langle$ecc$\rangle$}\\
\colhead{} & \colhead{} & \colhead{($\sigma_{\langle\text{v}_{\text{r}}\rangle}$,$\sigma_{\langle\text{v}_{\phi}\rangle}$,$\sigma_{\langle\text{v}_{\text{z}}\rangle}$)} & \colhead{($\sigma_{\langle\text{J}_{\text{r}}\rangle}$,$\sigma_{\langle\text{J}_{\phi}\rangle}$,$\sigma_{\langle\text{J}_{\text{z}}\rangle}$)} & \colhead{$\sigma_{\langle\text{E}\rangle}$} & \colhead{$\sigma_{\langle\text{ecc}\rangle}$}\\
\colhead{} & \colhead{} & \colhead{(km s$^{-1}$)} & \colhead{(kpc km s$^{-1}$)} & \colhead{(10$^{5}$ km$^{2}$ s$^{-2}$)} & \colhead{}}
\startdata
$DTG-1$ & $35$ & ($12.9$,$133.3$,$7.6$) & ($140.5$,$899.3$,$44.5$) & $-1.810$ & $0.419$ \\
 & & ($59.1$,$9.8$,$39.8$) & ($23.8$,$44.2$,$10.4$) &  \textcolor{white}{+}$0.018$ & $0.034$ \\
$DTG-2$ & $30$ & ($-8.7$,$259.0$,$0.7$) & ($23.8$,$2116.6$,$2.6$) & $-1.496$ & $0.126$ \\
 & & ($19.5$,$2.8$,$15.7$) & ($4.8$,$14.5$,$2.1$) &  \textcolor{white}{+}$0.003$ & $0.012$ \\
$DTG-3$ & $30$ & ($13.1$,$183.3$,$-3.7$) & ($46.7$,$1409.1$,$61.6$) & $-1.662$ & $0.210$ \\
 & & ($41.2$,$11.9$,$52.0$) & ($9.8$,$40.5$,$4.7$) &  \textcolor{white}{+}$0.012$ & $0.022$ \\
$DTG-4$ & $28$ & ($18.2$,$2.5$,$65.8$) & ($196.1$,$19.3$,$977.0$) & $-1.631$ & $0.475$ \\
 & & ($81.5$,$44.4$,$208.9$) & ($143.5$,$293.3$,$104.0$) &  \textcolor{white}{+}$0.053$ & $0.154$ \\
$DTG-5$ & $27$ & ($-2.5$,$189.7$,$-25.8$) & ($29.5$,$1392.6$,$105.7$) & $-1.655$ & $0.167$ \\
 & & ($35.7$,$16.0$,$67.3$) & ($11.4$,$53.0$,$6.0$) &  \textcolor{white}{+}$0.014$ & $0.035$ \\
$DTG-6$ & $24$ & ($-0.8$,$237.5$,$-1.6$) & ($3.4$,$1936.8$,$1.4$) & $-1.551$ & $0.050$ \\
 & & ($15.2$,$4.4$,$8.9$) & ($1.9$,$12.0$,$0.9$) &  \textcolor{white}{+}$0.004$ & $0.014$ \\
$DTG-7$ & $23$ & ($29.8$,$15.1$,$-2.3$) & ($573.1$,$107.5$,$31.5$) & $-1.845$ & $0.922$ \\
 & & ($75.8$,$16.6$,$36.6$) & ($28.8$,$121.9$,$7.3$) &  \textcolor{white}{+}$0.024$ & $0.044$ \\
$DTG-8$ & $23$ & ($1.2$,$2.8$,$3.6$) & ($826.9$,$24.6$,$38.5$) & $-1.708$ & $0.971$ \\
 & & ($144.0$,$9.5$,$30.3$) & ($32.8$,$78.7$,$21.0$) &  \textcolor{white}{+}$0.025$ & $0.033$ \\
$DTG-9$ & $23$ & ($13.2$,$204.9$,$0.0$) & ($21.3$,$1638.7$,$2.0$) & $-1.632$ & $0.134$ \\
 & & ($22.0$,$4.2$,$10.3$) & ($4.9$,$14.0$,$1.2$) &  \textcolor{white}{+}$0.003$ & $0.015$ \\
$DTG-10$ & $22$ & ($-5.4$,$187.5$,$3.5$) & ($45.2$,$1515.3$,$2.6$) & $-1.661$ & $0.199$ \\
 & & ($19.4$,$2.7$,$8.4$) & ($4.4$,$25.5$,$1.6$) &  \textcolor{white}{+}$0.009$ & $0.009$ \\
\enddata
\tablecomments{This table is a stub; the full table is available in the electronic edition.}\end{deluxetable*}

\begin{deluxetable*}{l  r  r  r  r  c  c  c  r}
\tablecaption{Identified Milky Way Substructures \label{tab:substructures}}
\tablehead{\colhead{MW Substructure} & \colhead{$N$ Stars} & \colhead{$\langle$[Fe/H]$\rangle$} & \colhead{$\langle$[C/Fe]$_{\textit{c}}\rangle$} & \colhead{$\langle$[$\alpha$/Fe]$\rangle$} & \colhead{($\langle$v$_{\text{r}}\rangle$,$\langle$v$_{\phi}\rangle$,$\langle$v$_{\text{z}}\rangle$)} & \colhead{($\langle$J$_{\text{r}}\rangle$,$\langle$J$_{\phi}\rangle$,$\langle$J$_{\text{z}}\rangle$)} & \colhead{$\langle$E$\rangle$} & \colhead{$\langle$ecc$\rangle$}\\
\colhead{} & \colhead{} & \colhead{} & \colhead{} & \colhead{} & \colhead{($\sigma_{\langle\text{v}_{\text{r}}\rangle}$,$\sigma_{\langle\text{v}_{\phi}\rangle}$,$\sigma_{\langle\text{v}_{\text{z}}\rangle}$)} & \colhead{($\sigma_{\langle\text{J}_{\text{r}}\rangle}$,$\sigma_{\langle\text{J}_{\phi}\rangle}$,$\sigma_{\langle\text{J}_{\text{z}}\rangle}$)} & \colhead{$\sigma_{\langle\text{E}\rangle}$} & \colhead{$\sigma_{\langle\text{ecc}\rangle}$}\\
\colhead{} & \colhead{} & \colhead{} & \colhead{} & \colhead{} & \colhead{(km s$^{-1}$)} & \colhead{(kpc km s$^{-1}$)} & \colhead{(10$^{5}$ km$^{2}$ s$^{-2}$)} & \colhead{}}
\startdata
GSE & 523 & $-1.556$ & $+0.309$ & $+0.315$ & ($-9.6$,$10.5$,$-1.9$) & ($726.9$,$77.7$,$129.9$) & $-1.694$ & $0.876$\\
 &  & $0.532$ & $0.442$ & $0.177$ & ($159.6$,$35.6$,$74.9$) & ($367.0$,$255.4$,$115.0$) & \textcolor{white}{+}$0.151$ & $0.089$\\
\hline
MWTD & 119 & $-1.386$ & $+0.294$ & $+0.355$ & ($1.8$,$128.4$,$-7.2$) & ($202.8$,$933.9$,$130.3$) & $-1.717$ & $0.459$\\
 &  & $0.597$ & $0.249$ & $0.162$ & ($74.6$,$36.4$,$75.7$) & ($88.0$,$222.3$,$95.1$) & \textcolor{white}{+}$0.063$ & $0.132$\\
\hline
Splashed Disk & 56 & $-1.109$ & $+0.478$ & $+0.516$ & ($-0.5$,$182.3$,$-2.6$) & ($58.0$,$1358.0$,$67.0$) & $-1.673$ & $0.228$\\
 &  & $0.413$ & $0.294$ & $0.113$ & ($38.6$,$19.5$,$49.1$) & ($30.9$,$65.4$,$47.3$) & \textcolor{white}{+}$0.021$ & $0.069$\\
\hline
Thamnos & 40 & $-2.004$ & $+0.252$ & $+0.364$ & ($12.0$,$-124.3$,$-10.5$) & ($197.4$,$-822.0$,$125.9$) & $-1.761$ & $0.482$\\
 &  & $0.541$ & $0.310$ & $0.139$ & ($75.0$,$30.5$,$67.5$) & ($78.0$,$193.4$,$86.5$) & \textcolor{white}{+}$0.044$ & $0.101$\\
\hline
Helmi Stream & 12 & $-1.950$ & $+0.192$ & $+0.360$ & ($-23.1$,$140.9$,$-139.9$) & ($366.8$,$1098.2$,$1149.4$) & $-1.342$ & $0.434$\\
 &  & $0.548$ & $0.255$ & $0.082$ & ($111.0$,$43.7$,$183.0$) & ($133.0$,$244.9$,$58.7$) & \textcolor{white}{+}$0.054$ & $0.065$\\
\hline
LMS-1 & 10 & $-2.461$ & $+0.366$ & $+0.349$ & ($62.5$,$49.1$,$3.9$) & ($595.7$,$346.7$,$1985.4$) & $-1.268$ & $0.514$\\
 &  & $0.411$ & $0.369$ & $0.173$ & ($134.3$,$42.9$,$279.7$) & ($201.8$,$286.7$,$73.0$) & \textcolor{white}{+}$0.055$ & $0.077$\\
\hline
\enddata
\end{deluxetable*}

\subsubsection{Gaia-Sausage-Enceladus}\label{subsubsec:GSE}

The most populated substructure is Gaia-Sausage-Enceladus (GSE), which contains $523$ member stars. GSE is thought to be a remnant of an earlier merger that distributed a significant number of stars throughout the inner halo of the MW \citep{Belokurov2018,Helmi2018}. The action space determined by the member stars exhibits an extended radial component, a null azimuthal component within errors, and a null vertical component. These orbital properties are the product of the high-eccentricity selection of the DTGs, and agree with previous findings of GSE orbital characteristics when using other selection criteria \citep{Koppelman2018,Myeong2018b,Limberg2021a}. 

The $\langle$[Fe/H]$\rangle$ of GSE found in our work is rather metal poor ([Fe/H] $\sim -1.5$), consistent with studies of its metallicity in dynamical groupings, even though our sample contains more metal-rich stars that could have been associated with GSE \citep{Gudin2021,Limberg2021a}. The stars that form DTGs in GSE tend to favor the more metal-poor tail of the substructure, which is also seen in previous DTG analysis. The $\langle$[$\alpha$/Fe]$\rangle$ of GSE exhibits a relatively low level, consistent with the low-Mg structure detected by \citet{Hayes2018} and with Mg levels consistent with accreted structures simulated by \citet{Mackereth2019}. The $\alpha$-element enhancement seen in GSE is due to accretion of older stellar populations, consistent with known element abundance patterns. We also obtain a $\langle$[C/Fe]$_{\textit{c}}\rangle$ for GSE (note that stars from \citealt{Huang2021c} do not have measurements of [C/Fe]$_{\textit{c}}$ or [$\alpha$/Fe]). Finally, we can associate the globular clusters Ryu 879 (RLGC2), IC 1257, NGC 4833, NGC 5139 ($\omega$Cen), and NGC 6284 with GSE, based on DTGs with similar orbital characteristics of these globular clusters (see Sec. \ref{subsec:GCDG} for details). Note in Figure~\ref{fig:energy_actions} how GSE occupies a large region of the Lindblad diagram, concentrated in the planar and radial portions of the projected-action plot.

\subsubsection{The Metal-Weak Thick Disk}\label{subsubsec:MWTD}

The second-most populated substructure is the Metal-Weak Thick Disk (MWTD), which contains $119$ member stars. The MWTD is thought to have formed from either a merger scenario, possibly related to GSE, or the result of old stars born within the Solar radius migrating out to the Solar position due to tidal instabilities within the MW \citep{Carollo2019}. The non-existent radial and vertical velocity components, as well as the large positive azimuthal velocity component of the MWTD are all consistent with the velocity distribution for the MWTD from Carollo et al., even with the [Fe/H] cut containing more metal-rich stars than in their sample. The mean eccentricity distribution found within this substructure is also similar to that reported by Carollo et al., showing that the MWTD is a distinct component from the canonical thick disk (TD). Recently, both \citet{An2020} and \citet{Dietz2021} have presented evidence that the MWTD is an independent structure from the TD. The distribution in $\langle$[Fe/H]$\rangle$ and mean velocity space represents a stellar population consistent with the high-Mg population \citep{Hayes2018}, with the mean $\alpha$-element abundance being similar within errors. The $\langle$[C/Fe]$_{\textit{c}}\rangle$ abundance is also given for the MWTD, and shows an enhancement in carbon, possibly pointing to a relation with the strongly prograde CEMP structure found in \citet{Dietz2021}, which was attributed to the MWTD population.  Notice in Figure~\ref{fig:energy_actions} how the MWTD occupies a higher energy component of the disk (the gray dots mostly positioned with prograde orbits) in the Lindblad diagram.

\begin{deluxetable*}{l  l  l  c}
\tablecaption{Associations of Identified DTGs \label{tab:interesting_substructure}}
\tablehead{\colhead{Structure} & \colhead{Reference} & \colhead{Associations} & \colhead{Identified DTGs}}
\startdata
\multirow{8}{*}{MW Substructure} & \multirow{8}{*}{\citet{Naidu2020}} & \multirow{4}{*}{GSE} & 7, 8, 12, 17, 18, 20, 24, 26, 27, 28, 30, 44, 45, 46, 50\\
 &  &  & 56, 57, 58, 66, 69, 70, 71, 75, 76, 78, 79, 80, 92, 95, 96\\
 &  &  & 97, 98, 100, 111, 112, 113, 114, 115, 116, 133, 134, 135\\
 &  &  & 136, 137, 141, 142, 146, 156, 164, 166, 167, 168, 169\\ \cline{3-4}
 &  & MWTD & 14, 22, 29, 37, 43, 65, 90, 106, 107, 110, 132\\ \cline{3-4}
 &  & Thamnos & 47, 52, 102, 138, 170\\ \cline{3-4}
 &  & Splashed Disk & 5, 15, 86\\ \cline{3-4}
 &  & Helmi Stream & 42\\ \cline{3-4}
 &  & LMS-1 & 67\\ \cline{1-4}
\multirow{8}{*}{Globular Clusters} & \multirow{8}{*}{\citet{Vasiliev2021}} & Ryu~879~(RLGC~2) & 12, 57, 58\\ \cline{3-4}
 &  & IC~1257 & 66\\ \cline{3-4}
 &  & NGC~4833 & 7\\ \cline{3-4}
 &  & NGC~5139~($\omega$Cen) & 18\\ \cline{3-4}
 &  & NGC~6284 & 141\\ \cline{3-4}
 &  & NGC~6356 & 52\\ \cline{3-4}
 &  & NGC~6397 & 55\\ \cline{3-4}
 &  & NGC~6752 & 33\\ \cline{1-4}
\enddata
\end{deluxetable*}

\subsubsection{The Splashed Disk}\label{subsubsec:SD}

The third-most populated substructure is the Splashed Disk (SD), which contains $56$ member stars. The SD is thought to be a component of the primordial MW disk that was kinematically heated during the GSE merger event \citep{Helmi2018,DiMatteo2019,Belokurov2020}. The mean velocity components of the SD are consistent with a null radial and vertical velocity, while showing a large positive azimuthal velocity consistent with disk-like stars. The mean eccentricity of these stars is consistent with disk-like orbits. The SD consists of the most metal-rich substructure identified here. The high $\langle$[$\alpha$/Fe]$\rangle$ abundances for the SD shows that these stars are old, and they could be the result of a possible merger event, such as the one that created GSE. The $\langle$[C/Fe]$_{\textit{c}}\rangle$ abundance for the SD are high, which is in conjunction with the high mean $\alpha$-element abundances. Notice in Figure~\ref{fig:energy_actions} how the SD overlaps with the MWTD. This is due to the selection criteria only using metallicity and $\alpha$-element abundances to determine the SD stars \citep{Naidu2020}. Considering the SD is thought to be composed of stars that have been heated due to the GSE merger event, the positions of the SD stars in the Lindblad diagram does not show a relatively large deviation from disk-like orbits. More associations with the SD are needed to make any definitive claims.

\subsubsection{Thamnos}\label{subsubsec:Thamnos}

The fourth-most populated substructure is Thamnos, which contains $40$ member stars. Thamnos was proposed by \citet{Koppelman2019a} as a merger event that populated these stars in a retrograde orbit that is similar to TD stars. The low energy and strong retrograde rotation suggest that Thamnos merged with the MW long ago \citep{Koppelman2019a}. Here we find a similar low mean orbital energy and strong mean retrograde motion, and we recover as strong a retrograde motion as in \citet{Koppelman2019a}, within errors. The low mean metallicity, consistent with the value reported by \citet{Limberg2021a}, and elevated $\langle$[C/Fe]$_{c}\rangle$ of these stars also supports the merger being ancient. The $\langle$[$\alpha$/Fe]$\rangle$ is high, also suggesting an old population, consistent with \citet{Kordopatis2020}. Notice in Figure~\ref{fig:energy_actions} how Thamnos occupies a space that could be described as a retrograde version of disk stars.  

\subsubsection{The Helmi Stream}\label{subsubsec:Helmi}

The second-to-least populated substructure is the Helmi Stream (HS), which contains $12$ member stars. The HS is one of the first detected dynamical substructures in the MW using integral of motions \citep{Helmi1999}. The HS has a characteristically high vertical velocity, which separates it from other stars that lie in the disk, and can be seen in the sample here. The large uncertainty on vertical velocity of the HS members corresponds to the positive and negative vertical velocity components of the stream, with the negative vertical velocity population dominating, consistent with the members determined here \citep{Helmi2020}. The $\langle$[Fe/H]$\rangle$ of the HS is more metal-poor in this sample, compared to the known HS members ([Fe/H] $\sim -1.5$; \citealt{Koppelman2019b}). Recently however, \citet{Limberg2021b} noted that the metallicity range of HS is more metal-poor than previously expected, with stars reaching down to [Fe/H] $\sim -2.5$, which is consistent with the results presented here. Notice in Figure~\ref{fig:energy_actions} how the HS occupies a relatively isolated space in the Lindblad diagram, thanks to the large vertical velocity of the stars providing the extra energy compared to the other disk stars.

\subsubsection{LMS-1 (Wukong)}\label{subsubsec:LMS1}

The least populated substructure is LMS-1, which contains $10$ member stars. LMS-1 was first identified by \citet{Yuan2020a}, and was also detected by \citet{Naidu2020}, who called it Wukong. This structure is similar to GSE in terms of the velocity component, but is characterized by a higher energy along with a more metal-poor population \citep{Naidu2020}, which is shown in the small number of stars representing LMS-1 in our sample. Notice in Figure~\ref{fig:energy_actions} how LMS-1 overlaps with GSE in the Linblad Diagram, but these stars exhibit a lower eccentricity ($ecc \sim 0.35$) compared to the GSE stars ($ecc > 0.7$) forming their own distinct substructure, with a more clear division in the projected-action plot. 

\subsection{Previously Identified Dynamically Tagged Groups and Stellar Associations}\label{subsec:Prev_DTGs_Stellar_assoc}

Separately, we can compare the newly identified DTGs in this work with other dynamical groups identified by previous authors.  We take the mean group properties used to detect the previously identified groups and compare them to the mean and dispersion for the dynamical parameters of our identified DTGs. Stellar associations are also considered, allowing the identification of stars in our sample that belong to previously identified groups. For details on the previous work used in this process, see Paper I. The resulting dynamical associations between our identified DTGs and previously identified groups (along with substructure and globular cluster associations, see Section \ref{subsec:GCDG}) are listed in Table~\ref{tab:interesting_substructure}. Table~\ref{tab:cluster_results_stub} lists the individual stellar associations for each of our DTGs.

One example of associations of identified DTGs with past groups is DTG-42. This DTG was associated with the Helmi Stream through the procedure outlined in \citet{Naidu2020} (see Section~\ref{subsec:MWSubstructure} for more details). There were four stars associated with this DTG through a $5 \arcsec$ radius search of the DTG member stars. Two of the stars belong in HK18:Green, and the other two belong in NB20:H99, both of which are associated with the Helmi Stream \citep{Koppelman2018}.  DTG-42 is also dynamically associated with GM18a:S2, GM18b:S2, DG21:CDTG-15, and GL20:DTG-3, all of which were identified as part of the Helmi Stream \citep{Myeong2018b,Myeong2018c,Gudin2021,Limberg2021a}. DG21:CDTG-15 was originally associated with GL20:DTG-3 as well by the authors, further strengthening our associations \citep{Gudin2021}. 

An interesting DTG associated with the GSE is DTG-7, which has multiple previously identified groups and stars associated with them. Taking a closer look at DTG-7, we can find three stellar associations between this group and DG21:CDTG-1, and one stellar association each in IR18:E and EV21:NGC 4833, a globular cluster \citep{Gudin2021,Roederer2018a,Vasiliev2021}. Again DG21:CDTG-1 is found associated with IR18:E by the authors, and even though we have a stellar association with EV21:NGC 4833, we do not find a strong dynamical association between DTG-7 and the globular cluster \citep{Gudin2021,Vasiliev2021}. DTG-7 is dynamically associated with GC21:Sausage, DG21:CDTG-1, GL21:DTG-37, DS21:DTG-43, again all of which are associated to GSE by their authors \citep{Cordoni2021,Gudin2021,Limberg2021a,Shank2021}. 

The only DTG with multiple associations related to the MWTD is DTG-14. Stellar associations with DG21:CDTG-6 and HL19:GL-1 were found, with neither of them having associations to large-scale substructure by their authors \citep{Gudin2021,Li2019}. Dynamical associations with both HL19:GL-1 and DS21:DTG-2 are found with DTG-14. Interestingly, since we use the same procedure outlined by \citet{Shank2021}, those authors found that DS21:DTG-2 is also associated with both HL19:GL-1 and DG21:CDTG-6 along with the MWTD. These associations across multiple papers show how powerful these identifications are when identifying past structures, such as the unidentified DG21:CTDG-6 and HL19:GL-1 as candidate MWTD associations. 

DTG-102 is an interesting case, since we have associated it with Thamnos, and there are 3 stellar associations with previously identified groups along with one dynamical association. There were two stellar associations between DTG-102 and HL20:GR-2, which was not identified by the authors \citep{Li2020}. DG21:CDTG-2 had a stellar association and was also dynamically associated with DTG-102 while being identified as a part of Thamnos by the authors \citep{Gudin2021}. 

Finally, there is an unassigned MW substructure DTG that has multiple associations in DTG-4. DTG-4 has four stellar associations, two of which are to DG21:CDTG-18 and the other two are associated with DS21:DTG-4. Both of these are unassigned by the authors, but DS21:DTG-4 was found to have associations with GC21:Sausage, DG21:CDTG-14, and GL21:DTG-13, of which GC21:Sausage is tentatively associated to GSE, and DG21:CDTG-14 is unassigned, while GL21:DTG-13 is associated to ZY20:DTG-39 which is unassigned \citep{Cordoni2021,Gudin2021,Limberg2021a,Yuan2020b}. Dynamical associations with DTG-4 include GM18a:C1, HL19:GL-2, GL21:DTG-8, and DS21:DTG-4. Of these associated groups, GL21:DTG-8 and DS21:DTG-4 have previous associations by thier authors. DS21:DTG-4 was already discussed in detail, while GL21:DTG-8 is previously associated to ZY20:DTG-35, which is also unassigned by their authors \citep{Myeong2018b,Li2019,Limberg2021a,Shank2021}. The DTGs that have unassigned MW substructure can offer valuable insights into the smaller structures of the MW that have yet to be confirmed.

Another use of the stellar associations comes from the suggestion by \citet{Roederer2018a}, strengthened by \citet{Gudin2021}, that dynamical groups of stars have a statistically significant correlation between their elemental abundances. This is of importance to discover new chemically peculiar stars, particularly  $r$-process-enhanced stars. Out of our DTGs, there are $22$ associations between known $r$-process-enhanced stars and our discovered DTGs. The member stars in these DTGs provide interesting candidates for high-resolution spectroscopic follow-up, due to the increased likelihood of the other members comprising chemically peculiar stars, especially in terms of $r$-process enhancement \citep{Roederer2018a,Gudin2021}. 

\subsection{Globular Clusters and Dwarf Galaxies}\label{subsec:GCDG}

Both globular clusters and dwarf galaxies have been shown to play an important role in the formation of chemically peculiar stars \citep{Ji2016a,Myeong2018a}. Globular clusters can also be a good indicator of 
galaxy-formation history based on their metallicities and orbits \citep{Woody2021}. From the work of \citet{Vasiliev2021}, we can compare the dynamical properties of $170$ globular clusters to those of the DTGs we identify. The procedure that is employed is the same one used for previously identified groups and stellar associations introduced in Sec. \ref{subsec:Prev_DTGs_Stellar_assoc}. The dynamics for $45$ dwarf galaxies of the MW (excluding the Large Magellanic Cloud, Small Magellanic Cloud, and Sagittarius) are also explored. Paper I contains details of the orbits of the globular clusters and dwarf galaxies. The same procedure used for previously identified groups was then applied to determine whether a DTG was dynamically associated to the dwarf galaxy. Stellar associations were also determined for both globular clusters and dwarf galaxies in the same manner as previously identified groups. 

The above comparison exercise led to $10$ globular cluster associations, with $8$ being unique. For a breakdown of which globular clusters are associated with our DTGs, see Table~\ref{tab:interesting_substructure}. Ryu 879 (RLGC 2) has three DTG associations which agree with each other in mean metallicity ($\langle$[Fe/H]$\rangle$ $\sim -1.7$). The mean metallicity agrees with the discovery of the globular cluster Ryu 879 (RLGC 2) within errors \citep{Ryu2018}.The rest of the globular cluster associations are each associated with only one DTG in this work. We identify IC 1257, NGC 6284, and NGC 6356 through dynamical association, while NGC 4833, NGC 5139 ($\omega$Cen), NGC 6397, and NGC 6752 are identified through stellar association. Even though the matched stars in these globular clusters would have individually been associated with the globular cluster orbital parameters, the overall DTG did not possess sufficiently similar orbital characteristics to be associated. 

\startlongtable
\begin{deluxetable*}{l  l  c}
\tablecaption{Associations of Identified DTGs with Previous Groups \label{tab:interesting_groups_substructure}}
\tablehead{\colhead{Reference} & \colhead{Associations} & \colhead{Identified DTGs}}
\startdata
\multirow{23}{*}{\citet{Shank2021}} & DTG-1 & 119, 147, 148\\ \cline{2-3}
 & DTG-2 & 14, 53\\ \cline{2-3}
 & DTG-17 & 17, 27\\ \cline{2-3}
 & DTG-29 & 20, 135\\ \cline{2-3}
 & DTG-3 & 8\\ \cline{2-3}
 & DTG-4 & 4\\ \cline{2-3}
 & DTG-5 & 26\\ \cline{2-3}
 & DTG-8 & 75\\ \cline{2-3}
 & DTG-9 & 55\\ \cline{2-3}
 & DTG-11 & 57\\ \cline{2-3}
 & DTG-12 & 10\\ \cline{2-3}
 & DTG-13 & 2\\ \cline{2-3}
 & DTG-14 & 163\\ \cline{2-3}
 & DTG-19 & 103\\ \cline{2-3}
 & DTG-22 & 133\\ \cline{2-3}
 & DTG-23 & 19\\ \cline{2-3}
 & DTG-27 & 28\\ \cline{2-3}
 & DTG-30 & 31\\ \cline{2-3}
 & DTG-36 & 95\\ \cline{2-3}
 & DTG-42 & 92\\ \cline{2-3}
 & DTG-43 & 7\\ \cline{2-3}
 & DTG-44 & 90\\ \cline{2-3}
 & DTG-47 & 6\\ \cline{1-3}
\multirow{17}{*}{\citet{Gudin2021}} & CDTG-5 & 15, 176\\ \cline{2-3}
 & CDTG-7 & 79, 141\\ \cline{2-3}
 & CDTG-22 & 70, 164\\ \cline{2-3}
 & CDTG-23 & 25, 124\\ \cline{2-3}
 & CDTG-1 & 7\\ \cline{2-3}
 & CDTG-2 & 102\\ \cline{2-3}
 & CDTG-3 & 51\\ \cline{2-3}
 & CDTG-4 & 67\\ \cline{2-3}
 & CDTG-6 & 14\\ \cline{2-3}
 & CDTG-9 & 8\\ \cline{2-3}
 & CDTG-12 & 111\\ \cline{2-3}
 & CDTG-13 & 46\\ \cline{2-3}
 & CDTG-14 & 78\\ \cline{2-3}
 & CDTG-15 & 42\\ \cline{2-3}
 & CDTG-18 & 4\\ \cline{2-3}
 & CDTG-19 & 20\\ \cline{2-3}
 & CDTG-28 & 30\\ \cline{1-3}
\multirow{7}{*}{\citet{Limberg2021a}} & DTG-37 & 7, 17, 71\\ \cline{2-3}
 & DTG-2 & 67\\ \cline{2-3}
 & DTG-3 & 42\\ \cline{2-3}
 & DTG-8 & 4\\ \cline{2-3}
 & DTG-11 & 92\\ \cline{2-3}
 & DTG-25 & 170\\ \cline{2-3}
 & DTG-34 & 8\\ \cline{1-3}
 \pagebreak
\multirow{6}{*}{\citet{Helmi2017}} & VelHel-6 & 26, 29, 44, 69, 70, 71, 93, 133, 166\\ \cline{2-3}
 & VelHel-7 & 94, 124, 132\\ \cline{2-3}
 & VelHel-2 & 98, 156\\ \cline{2-3}
 & VelHel-5 & 30, 46\\ \cline{2-3}
 & VelHel-1 & 98\\ \cline{2-3}
 & VelHel-4 & 47\\ \cline{1-3}
 \multirow{2}{*}{\citet{Borsato2020}} & Enc & 50\\ \cline{2-3}
 & H99 & 42\\ \cline{1-3}
\multirow{3}{*}{\citet{Cordoni2021}} & \multirow{2}{*}{Sausage} & 7, 8, 17, 24, 28, 30, 45, 50, 57, 58, 66, 75\\
 &  & 79, 80, 92, 98, 115, 134, 135, 136, 142, 146, 168\\ \cline{2-3}
 & Sequoia & 47, 52, 127, 138\\ \cline{1-3}
 \multirow{2}{*}{\citet{Li2019}} & GL-1 & 14, 38, 99\\ \cline{2-3}
 & GL-2 & 4\\ \cline{1-3}
\multirow{2}{*}{\citet{Li2020}} & GR-1 & 20, 27, 97, 114, 141\\ \cline{2-3}
 & GR-2 & 102, 138\\ \cline{1-3}
\multirow{2}{*}{\citet{Monty2020}} & Sausage & 28, 50, 66, 75, 92\\ \cline{2-3}
 & SeqG1 & 170\\ \cline{1-3}
\multirow{2}{*}{\citet{Myeong2018b}} & C1 & 4, 67\\ \cline{2-3}
 & S2 & 42\\ \cline{1-3}
\multirow{2}{*}{\citet{Myeong2018c}} & Cand11 & 67\\ \cline{2-3}
 & S2 & 42\\ \cline{1-3}
\multirow{1}{*}{\citet{Koppelman2018}} & Green & 42\\ \cline{1-3}
\multirow{1}{*}{\citet{Myeong2017}} & Comoving & 12, 58, 67, 75, 79, 80\\ \cline{1-3}
\multirow{1}{*}{\citet{Roederer2018a}} & E & 7\\
\enddata
\tablecomments{We draw attention to the associations with CDTGs from \citet{Gudin2021} and \citet{Roederer2018a} due to their enhancement in $r$-process abundances.}
\end{deluxetable*}

The DTGs associated with globular clusters are expected to have formed in chemically similar birth environments; this is mostly supported through the similar chemical properties of the DTGs. Associations of globular clusters with Galactic substructure have been made by \cite{Massari2019}. These authors did not analyze Ryu 879 (RLGC 2), since the globular cluster was only recently discovered at the time of the publication. IC 1257, NGC 6284, and NGC 4833 were identified to GSE by both our identification and \cite{Massari2019} (see DTG-66, DTG-141, and DTG-7 respectively). NGC 6356 was identified as Thamnos by our determinations, while \citet{Massari2019} found it to be associated to the main disk which we do not consider as part of the substructure routine, though it should be pointed out that Thamnos has the dynamics of a higher-energy retrograde disk similar to the MWTD \citep{Koppelman2019a}. NGC 5139 ($\omega$ Cen) was identified as being associated with GSE or Sequoia by \citet{Massari2019}, and our associations did recover this match to GSE (see DTG-18). NGC 6397 and NGC 6752 are not associated to any substructure by our procedure (see DTG-55 and DTG-33, respectively), but \cite{Massari2019} find them associated to the main disk, which is not a part of the substructure routine. 

We did not identify any associations of DTGs to the sample of (surviving) MW dwarf galaxies, either through stellar associations, or through the dynamical association procedure described above. Nevertheless, some of the DTGs identified by our analysis may well be associated with dwarf galaxies that have previously merged with the MW.

The analysis on the global properties of the identified DTGs that was explored in Section 6 of Paper I is foregone in this analysis due to the small number of DTGs that have sufficient C and $\alpha$-element abundances to derive meaningful conclusions. Attention to the small number of DTGs that do meet these requirements are listed in Table~\ref{tab:cluster_results_element_stub}.

\section{Discussion}\label{sec:Discussion}

We have assembled a Full Sample of $8675$ stars from the RAVE DR6 survey \citep{Steinmetz2020a} with available estimates of [Fe/H], and in some cases, with [C/Fe] and [$\alpha$/Fe]. 

The Initial Sample contained $8377$ stars, with $4250 \leq $ T$_{eff}$ (K) $\leq 7000$ and [Fe/H] $\leq -0.8$, $106$ of which we identify as CEMP stars ([C/Fe]$_c >$ +0.7); these are listed in Table~\ref{tab:cemp} in the Appendix. These stars are of interest due to their enhanced carbon and association to morphological groups described in \citet{Yoon2016}. Based on their classification scheme, there are approximately $50$ Group I CEMP stars, $55$ Group II CEMP stars, and $5$ Group III CEMP stars, with a number of stars that have ambiguous classifications. This list provides a useful reference for high-resolution follow-up targets, some of which has already begun (e.g., \citealt{Rasmussen2020} and \citealt{Zepeda2021}).

The Final Sample of $7957$ metal-poor stars had sufficient radial velocity and astrometric information from which orbits were constructed, in order to determine Dynamically Tagged Groups (DTGs) in orbital energy and cylindrical action space with the \HDBSCAN ~algorithm. We chose \HDBSCAN\ as the clustering algorithm due to precedence within the literature \citep{Koppelman2019a,Gudin2021,Limberg2021a,Shank2021}, and its ability to extract clusters of stars over the energy and action space. Other clustering algorithms have been considered in the past, such as agglomerative clustering, affinity propogation, K-means, and mean-shift clustering \citep{Roederer2018a}, along with friends-of-friends \citep{Gudin2021}. 

We recover $179$ DTGs that include between $5$ and $35$ members, with $67$ DTGs containing at least $10$ member stars. These DTGs were associated with MW substructure, resulting in the identification of the Gaia-Sausage-Enceladus, the Metal-Weak Thick Disk, the Splashed Disk, Thamnos, the Helmi Stream, and LMS-1 (Wukong). A total of $8$ unique globular clusters were associated with $10$ different DTGs, while no surviving dwarf galaxies were determined to be associated with the identified DTGs. Previously identified groups were found to be associated with the DTGs as well, with past work mostly confirming our substructure identification. Each of these associations allow insights into the dynamical and chemical properties of the parent substructures. 

The implications of past group and stellar associations were explored with emphasis placed on the structure associations. Chemically peculiar stellar associations and previously identified Chemo-Dynamically Tagged Groups (CDTGs) were addressed as being good candidates for high-resolution follow-up spectroscopy targets, due to the statistical likelihood of the other members being chemically peculiar as well, mostly focused on $r$-process-enhanced stars. 

The methods presented here will be used on larger samples of field stars that we are in the process of assembling -- the HK/HES/HKII surveys \citep{Beers1985,Beers1992,Christlieb2008,Rhee2001} (a subset of which were analyzed by \citealt{Limberg2021a}), as well as SDSS/LAMOST and APOGEE. These data sets will also be supplemented with photometric estimates of effective temperature and metallicity from \citet{Huang2021c}, which allows stars from the HK/HES/HKII surveys with no previous spectroscopic follow-up to be explored, expanding the data set used by \citet{Limberg2021a}. 

\vspace{2.0cm}

\section{Appendix}

Here we present the tables for the Initial (Table~\ref{tab:initial_data_descript}) and Final (Table~\ref{tab:final_data_descript}) Samples of the RAVE DR6 survey. We also present Table~\ref{tab:cemp} which describes the identified CEMP stars and their associated morphological groups, according to the regions defined by \citet{Yoon2016}.

\startlongtable
\begin{deluxetable*}{c  l  l  l}
\tablecaption{Description of the Initial Sample from the RAVE Survey \label{tab:initial_data_descript}}
\tablehead{\colhead{Column} & \colhead{Field} & \colhead{Unit} & \colhead{Description}}
\startdata
1 & RAVE ID & $-$ & The RAVE DR6 RAVE ID of the star\\
2 & Source ID & $-$ & The Gaia EDR3 Source ID of the star\\
3 & SMSS ID & $-$ & The SMSS DR2 Source ID of the star\\
4 & RAVE NAME & $-$ & The RAVE DR6 RAVE NAME of the star\\
5 & RA & (J2000) & The Right Ascension of the star given in hours:minutes:seconds\\
6 & DEC & (J2000) & The Declination of the star given in degrees:minutes:seconds\\
7 & Telescope & $-$ & The telescope that was used to obtain the spectrum of the star\\
8 & $V_{\rm mag}$ & $-$ & The $V$ magnitude of the star as given by the $V_{\rm mag}$ Reference\\
9 & $G_{\rm mag}$ & $-$ & The Gaia $G$ mean magnitude of the star as given by the Gaia Source ID\\
10 & $G_{\rm BP} - G_{\rm RP}$ & $-$ & The Gaia BP $-$ RP color mean magnitude of the star as given by the Gaia Source ID\\
11 & $V_{\rm mag}$ (Gaia) & $-$ & The $V$ magnitude of the star as determined by the transformations from G mag to V mag using\\
 &  &  & $V$ $=$ $G$ $+$ $0.02704$ $-$ $0.01424*(\rm{BP}-\rm{RP})$ $+$ $0.2156*(\rm{BP}-\rm{RP})^2$ $-$\\
 &  &  & $0.01426(\rm{BP}-\rm{RP})^3$ given by \citet{Riello2021}\\
12 & RV$_{\rm RAVE}$ & (km s$^{-1}$) & The radial velocity as given by the RAVE ID\\
13 & Error & (km s$^{-1}$) & The radial velocity error as given by the RAVE ID\\
14 & RV$_{\rm Gaia}$ & (km s$^{-1}$) & The radial velocity as given by the Gaia Source ID\\
15 & Error & (km s$^{-1}$) & The radial velocity error as given by the Gaia Source ID\\
16 & Parallax & (mas) & The parallax as given by the Gaia Source ID\\
17 & Error & (mas) & The parallax error as given by the Gaia Source ID\\
18 & Distance & (kpc) & The inverse parallax distance (1/Parallax)\\
19 & Error & (kpc) & The inverse parallax distance error (Parallax$_{error}$/(Parallax$^2$))\\
20 & Distance$_{\rm Corrected}$ & (kpc) & The corrected inverse parallax distance (1/(Parallax + 0.026)) based on \citet{Huang2021a}\\
21 & Error & (kpc) & The corrected inverse parallax distance error (Parallax$_{error}$/((Parallax + 0.026)$^2$))\\
 &  &  & based on \citet{Huang2021a}\\
22 & Relative Error & $-$ & The relative error of the corrected distance as given by Gaia\\
23 & Distance BJ21 & (kpc) & The 50 percentile distance as given by \citet{Bailer-Jones2021} based on the Gaia Source ID\\
24 & Error & (kpc) & The 50 percentile error as estimated by the 84 percentile distance and the 16 percentile\\
 &  &  & distance as given by \citet{Bailer-Jones2021} based on the Gaia Source ID ((dist84-dist16)/2)\\
25 & Relative Error & $-$ & The relative error of the 50 percentile distance as given by \citet{Bailer-Jones2021} based on\\
 &  &  & the Gaia Source ID\\
26 & Distance StarHorse & (kpc) & The 50 percentile distance as given by \citet{Anders2021} based on the Gaia Source ID\\
27 & Error & (kpc) & The 50 percentile error as estimated by the 84 percentile distance and the 16 percentile\\
 &  &  & distance as given by \citet{Anders2021} based on the Gaia Source ID ((dist84-dist16)/2)\\
28 & Relative Error & $-$ & The relative error of the 50 percentile distance as given by \citet{Anders2021} based on the\\
 &  &  & Gaia Source ID\\
29 & PM$_{\rm RA}$ & (mas yr$^{-1}$) & The proper motion in the Right Ascension as given by the Gaia Source ID\\
30 & Error & (mas yr$^{-1}$) & The proper motion error in the Right Ascension as given by the Gaia Source ID\\
31 & PM$_{\rm DEC}$ & (mas yr$^{-1}$) & The proper motion in the Declination as given by the Gaia Source ID\\
32 & Error & (mas yr$^{-1}$) & The proper motion error in the Declination as given by the Gaia Source ID\\
33 & Correlation Coefficient & $-$ & The correlation coefficient between the proper motion in Right Ascension and the proper motion\\
 &  &  & in Declination as given by the Gaia Source ID\\
34 & T$_{\rm eff~RAVE}$ & (K) & The effective temperature of the star as given by the RAVE ID\\
35 & Error & (K) & The effective temperature error of the star as given by the RAVE ID\\
36 & T$_{\rm eff~Spec}$ & (K) & The effective temperature of the star as given by the n-SSPP\\
37 & Error & (K) & The effective temperature error of the star as given by the n-SSPP\\
38 & T$_{\rm eff~Phot}$ & (K) & The effective temperature of the star as given by \citet{Huang2021c}\\
39 & Error & (K) & The effective temperature error of the star as given by \citet{Huang2021c}\\
40 & T$_{\rm eff}$ & (K) & The adopted effective temperature of the star based on the Parameter Procedure\\
41 & Error & (K) & The adopted effective temperature error of the star based on the Parameter Procedure\\
42 & log \textit{g}$_{\rm RAVE}$ & (cgs) & The surface gravity of the star as given by the RAVE ID\\
43 & Error & (cgs) & The surface gravity error of the star as given by the RAVE ID\\
44 & log \textit{g}$_{\rm Spec}$ & (cgs) & The surface gravity of the star as given by the n-SSPP\\
\pagebreak \\[-3.5ex]
45 & Error & (cgs) & The surface gravity error of the star as given by the n-SSPP\\
46 & log \textit{g} & (cgs) & The adopted surface gravity of the star based on the Parameter Procedure\\
47 & Error & (cgs) & The adopted surface gravity error of the star based on the Parameter Procedure\\
48 & [M/H] & $-$ & The metallicity of the star as given by the RAVE ID\\
49 & Error & $-$ & The metallicity error of the star as given by the RAVE ID\\
50 & [Fe/H]$_{Spec}$ & $-$ & The metallicity of the star as given by the n-SSPP\\
51 & Error & $-$ & The metallicity error of the star as given by the n-SSPP\\
52 & [Fe/H]$_{Phot}$ & $-$ & The metallicity of the star as given by \citet{Huang2021c}\\
53 & Error & $-$ & The metallicity error of the star as given by \citet{Huang2021c}\\
54 & [Fe/H] & $-$ & The adopted metallicity of the star based on the Parameter Procedure\\
55 & Error & $-$ & The adopted metallicity error of the star based on the Parameter Procedure\\
56 & [C/Fe] & $-$ & The carbon abundance ratio for the star\\
57 & Error & $-$ & The carbon abundance ratio error for the star\\
58 & [C/Fe]$_{c}$ & $-$ & The carbon abundance corrected for evolutionary effects from \citet{Placco2014}\\
59 & AC$_{c}$ & $-$ & The absolute carbon corrected for evolutionary effects from \citet{Placco2014} ([C/Fe]$_{c}$ +\\
 &  &  & [Fe/H] + log($\epsilon$)$_{Carbon,Solar}$) (Taken from solar value of 8.43 from\\
 &  &  & \citet{Asplund2009})\\
60 & CARDET & $-$ & Flag with ``D" if the carbon abundance ([C/Fe]) is detected from n-SSPP and ``U" if an upper\\
 &  &  & limit by n-SSPP and ``L" if a lower limit by n-SSPP and ``N" if none is detected by n-SSPP\\
61 & CC$_{\rm [C/Fe]}$ & $-$ & The correlation coefficient of [C/Fe] as given by the n-SSPP\\
62 & CEMP & $-$ & Flag as ``C" for Carbon-Enhanced Metal-Poor (CEMP) if [C/Fe]$_{c}$ $> +0.7$ and ``I" for\\
 &  &  & CEMP-intermediate if $+0.5 <$ [C/Fe]$_{c}$ $\leq +0.7$ and ``N" for Carbon-Normal if\\
 &  &  & [C/Fe]$_{c}$ $\leq +0.5$ and ``X" if there is no [C/Fe]$_{c}$ information\\
63 & [$\alpha$/Fe]$_{\rm RAVE}$ & $-$ & The alpha-element abundance ratio for the star as given by the RAVE ID\\
64 & Error & $-$ & The alpha-element abundance ratio error for the star as given by the RAVE ID\\
65 & [$\alpha$/Fe]$_{\rm Spec}$ & $-$ & The alpha-element abundance ratio for the star as given by the n-SSPP\\
66 & Error & $-$ & The alpha-element abundance ratio error for the star as given by the n-SSPP\\
67 & ALPDET & $-$ & Flag with ``D" if the alpha abundance ([$\alpha$/Fe]) is detected from n-SSPP and ``U" if an\\
 &  &  & upper limit by n-SSPP and ``L" if a lower limit by n-SSPP and ``N" if none is detected by\\
 &  &  & n-SSPP\\
68 & CC$_{\rm{[}\alpha\rm{/Fe]}}$ & $-$ & The correlation coefficient of [$\alpha$/Fe] as detected by the n-SSPP\\
69 & [$\alpha$/Fe] & $-$ & The alpha-element abundance ratio for the star based on the Parameter Procedure\\
70 & Error & $-$ & The alpha-element abundance ratio error for the star based on the Parameter Procedure\\
71 & SNR & $-$ & The average Signal-to-Noise Ratio of the spectrum from the RAVE ID\\
72 & Reference & $-$ & The Reference for the star as given by ``RAVE DR6" for the sample of stars from\\
 &  &  & \citet{Steinmetz2020a} sample and ``RAVE DR5" for the sample of stars from \citet{Kunder2017}\\
73 & Parameter Procedure & $-$ & The procedure used to determine the adopted stellar parameters (T$_{\rm eff}$, log \textit{g},\\
 &  &  & [Fe/H], and [$\alpha$/Fe]) (``Average" is used if the difference between [Fe/H]$_{Spec}$ and\\
 &  &  & [Fe/H]$_{Phot}$ is less $\leq$\ 0.5 dex and ``Spectroscopic" is used if only [Fe/H]$_{Spec}$\\
 &  &  & is available and ``Photometric" is used if only [Fe/H]$_{Phot}$ is available, while a choice\\
 &  &  & is made between [Fe/H]$_{Spec}$ and [Fe/H]$_{Phot}$ if both are available and the difference\\
 &  &  & is $>$ 0.5 dex)\\
74 & $V_{\rm mag}$ Reference & $-$ & The Reference for the \textit{V} magnitude of the star\\
75 & Distance AGAMA & $-$ & The Reference for the distance used in AGAMA (StarHorse prioritized over BJ21 unless StarHorse\\
 &  &  & distance has relative error greater than 0.3, if both have a relative error greater than 0.3\\
 &  &  & we adopt no distance estimate)\\
76 & RV Reference & $-$ & The Reference for the RV\\
\enddata
\end{deluxetable*}

\startlongtable
\begin{deluxetable*}{c  l  l  l}
\tablecaption{Description of the Final Sample from the RAVE Survey \label{tab:final_data_descript}}
\tablehead{\colhead{Column} & \colhead{Field} & \colhead{Unit} & \colhead{Description}}
\startdata
1 & RAVE ID & $-$ & The RAVE DR6 RAVE ID of the star\\
2 & Source ID & $-$ & The Gaia EDR3 Source ID of the star\\
3 & SMSS ID & $-$ & The SMSS DR2 Source ID of the star\\
4 & RAVE NAME & $-$ & The RAVE DR6 RAVE NAME of the star\\
5 & RA & (J2000) & The Right Ascension of the star given in hours:minutes:seconds\\
6 & DEC & (J2000) & The Declination of the star given in degrees:minutes:seconds\\
7 & Telescope & $-$ & The telescope that was used to obtain the spectrum of the star\\
8 & $V_{\rm mag}$ & $-$ & The $V$ magnitude of the star as given by the $V_{\rm mag}$ Reference\\
9 & $G_{\rm mag}$ & $-$ & The Gaia $G$ mean magnitude of the star as given by the Gaia Source ID\\
10 & $G_{\rm BP} - G_{\rm RP}$ & $-$ & The Gaia BP $-$ RP color mean magnitude of the star as given by the Gaia Source ID\\
11 & $V_{\rm mag}$ (Gaia) & $-$ & The $V$ magnitude of the star as determined by the transformations from G mag to V mag using\\
 &  &  & $V$ $=$ $G$ $+$ $0.02704$ $-$ $0.01424*(\rm{BP}-\rm{RP})$ $+$ $0.2156*(\rm{BP}-\rm{RP})^2$ $-$\\
 &  &  & $0.01426(\rm{BP}-\rm{RP})^3$ given by \citet{Riello2021}\\
12 & RV$_{\rm RAVE}$ & (km s$^{-1}$) & The radial velocity as given by the RAVE ID\\
13 & Error & (km s$^{-1}$) & The radial velocity error as given by the RAVE ID\\
14 & RV$_{\rm Gaia}$ & (km s$^{-1}$) & The radial velocity as given by the Gaia Source ID\\
15 & Error & (km s$^{-1}$) & The radial velocity error as given by the Gaia Source ID\\
16 & Parallax & (mas) & The parallax as given by the Gaia Source ID\\
17 & Error & (mas) & The parallax error as given by the Gaia Source ID\\
18 & Distance & (kpc) & The inverse parallax distance (1/Parallax)\\
19 & Error & (kpc) & The inverse parallax distance error (Parallax$_{error}$/(Parallax$^2$))\\
20 & Distance$_{\rm Corrected}$ & (kpc) & The corrected inverse parallax distance (1/(Parallax + 0.026)) based on \citet{Huang2021a}\\
21 & Error & (kpc) & The corrected inverse parallax distance error (Parallax$_{error}$/((Parallax + 0.026)$^2$))\\
 &  &  & based on \citet{Huang2021a}\\
22 & Relative Error & $-$ & The relative error of the corrected distance as given by Gaia\\
23 & Distance BJ21 & (kpc) & The 50 percentile distance as given by \citet{Bailer-Jones2021} based on the Gaia Source ID\\
24 & Error & (kpc) & The 50 percentile error as estimated by the 84 percentile distance and the 16 percentile\\
 &  &  & distance as given by \citet{Bailer-Jones2021} based on the Gaia Source ID ((dist84-dist16)/2)\\
25 & Relative Error & $-$ & The relative error of the 50 percentile distance as given by \citet{Bailer-Jones2021} based on\\
 &  &  & the Gaia Source ID\\
26 & Distance StarHorse & (kpc) & The 50 percentile distance as given by \citet{Anders2021} based on the Gaia Source ID\\
27 & Error & (kpc) & The 50 percentile error as estimated by the 84 percentile distance and the 16 percentile\\
 &  &  & distance as given by \citet{Anders2021} based on the Gaia Source ID ((dist84-dist16)/2)\\
28 & Relative Error & $-$ & The relative error of the 50 percentile distance as given by \citet{Anders2021} based on the\\
 &  &  & Gaia Source ID\\
29 & PM$_{\rm RA}$ & (mas yr$^{-1}$) & The proper motion in the Right Ascension as given by the Gaia Source ID\\
30 & Error & (mas yr$^{-1}$) & The proper motion error in the Right Ascension as given by the Gaia Source ID\\
31 & PM$_{\rm DEC}$ & (mas yr$^{-1}$) & The proper motion in the Declination as given by the Gaia Source ID\\
32 & Error & (mas yr$^{-1}$) & The proper motion error in the Declination as given by the Gaia Source ID\\
33 & Correlation Coefficient & $-$ & The correlation coefficient between the proper motion in Right Ascension and the proper motion\\
 &  &  & in Declination as given by the Gaia Source ID\\
34 & T$_{\rm eff~RAVE}$ & (K) & The effective temperature of the star as given by the RAVE ID\\
35 & Error & (K) & The effective temperature error of the star as given by the RAVE ID\\
36 & T$_{\rm eff~Spec}$ & (K) & The effective temperature of the star as given by the n-SSPP\\
37 & Error & (K) & The effective temperature error of the star as given by the n-SSPP\\
38 & T$_{\rm eff~Phot}$ & (K) & The effective temperature of the star as given by \citet{Huang2021c}\\
39 & Error & (K) & The effective temperature error of the star as given by \citet{Huang2021c}\\
40 & T$_{\rm eff}$ & (K) & The adopted effective temperature of the star based on the Parameter Procedure\\
41 & Error & (K) & The adopted effective temperature error of the star based on the Parameter Procedure\\
42 & log \textit{g}$_{\rm RAVE}$ & (cgs) & The surface gravity of the star as given by the RAVE ID\\
43 & Error & (cgs) & The surface gravity error of the star as given by the RAVE ID\\
44 & log \textit{g}$_{\rm Spec}$ & (cgs) & The surface gravity of the star as given by the n-SSPP\\
\pagebreak \\[-3.5ex]
45 & Error & (cgs) & The surface gravity error of the star as given by the n-SSPP\\
46 & log \textit{g} & (cgs) & The adopted surface gravity of the star based on the Parameter Procedure\\
47 & Error & (cgs) & The adopted surface gravity error of the star based on the Parameter Procedure\\
48 & [M/H] & $-$ & The metallicity of the star as given by the RAVE ID\\
49 & Error & $-$ & The metallicity error of the star as given by the RAVE ID\\
50 & [Fe/H]$_{Spec}$ & $-$ & The metallicity of the star as given by the n-SSPP\\
51 & Error & $-$ & The metallicity error of the star as given by the n-SSPP\\
52 & [Fe/H]$_{Phot}$ & $-$ & The metallicity of the star as given by \citet{Huang2021c}\\
53 & Error & $-$ & The metallicity error of the star as given by \citet{Huang2021c}\\
54 & [Fe/H] & $-$ & The adopted metallicity of the star based on the Parameter Procedure\\
55 & Error & $-$ & The adopted metallicity error of the star based on the Parameter Procedure\\
56 & [C/Fe] & $-$ & The carbon abundance ratio for the star\\
57 & Error & $-$ & The carbon abundance ratio error for the star\\
58 & [C/Fe]$_{c}$ & $-$ & The carbon abundance corrected for evolutionary effects from \citet{Placco2014}\\
59 & AC$_{c}$ & $-$ & The absolute carbon corrected for evolutionary effects from \citet{Placco2014} ([C/Fe]$_{c}$ +\\
 &  &  & [Fe/H] + log($\epsilon$)$_{Carbon,Solar}$) (Taken from solar value of 8.43 from\\
 &  &  & \citet{Asplund2009})\\
60 & CARDET & $-$ & Flag with ``D" if the carbon abundance ([C/Fe]) is detected from n-SSPP and ``U" if an upper\\
 &  &  & limit by n-SSPP and ``L" if a lower limit by n-SSPP and ``N" if none is detected by n-SSPP\\
61 & CC$_{\rm [C/Fe]}$ & $-$ & The correlation coefficient of [C/Fe] as given by the n-SSPP\\
62 & CEMP & $-$ & Flag as ``C" for Carbon-Enhanced Metal-Poor (CEMP) if [C/Fe]$_{c}$ $> +0.7$ and ``I" for\\
 &  &  & CEMP-intermediate if $+0.5 <$ [C/Fe]$_{c}$ $\leq +0.7$ and ``N" for Carbon-Normal if\\
 &  &  & [C/Fe]$_{c}$ $\leq +0.5$ and ``X" if there is no [C/Fe]$_{c}$ information\\
63 & [$\alpha$/Fe]$_{\rm RAVE}$ & $-$ & The alpha-element abundance ratio for the star as given by the RAVE ID\\
64 & Error & $-$ & The alpha-element abundance ratio error for the star as given by the RAVE ID\\
65 & [$\alpha$/Fe]$_{\rm Spec}$ & $-$ & The alpha-element abundance ratio for the star as given by the n-SSPP\\
66 & Error & $-$ & The alpha-element abundance ratio error for the star as given by the n-SSPP\\
67 & ALPDET & $-$ & Flag with ``D" if the alpha abundance ([$\alpha$/Fe]) is detected from n-SSPP and ``U" if an\\
 &  &  & upper limit by n-SSPP and ``L" if a lower limit by n-SSPP and ``N" if none is detected by\\
 &  &  & n-SSPP\\
68 & CC$_{\rm{[}\alpha\rm{/Fe]}}$ & $-$ & The correlation coefficient of [$\alpha$/Fe] as detected by the n-SSPP\\
69 & [$\alpha$/Fe] & $-$ & The alpha-element abundance ratio for the star based on the Parameter Procedure\\
70 & Error & $-$ & The alpha-element abundance ratio error for the star based on the Parameter Procedure\\
71 & SNR & $-$ & The average Signal-to-Noise Ratio of the spectrum from the RAVE ID\\
72 & Reference & $-$ & The Reference for the star as given by ``RAVE DR6" for the sample of stars from\\
 &  &  & \citet{Steinmetz2020a} sample and ``RAVE DR5" for the sample of stars from \citet{Kunder2017}\\
73 & Parameter Procedure & $-$ & The procedure used to determine the adopted stellar parameters (T$_{\rm eff}$, log \textit{g},\\
 &  &  & [Fe/H], and [$\alpha$/Fe]) (``Average" is used if the difference between [Fe/H]$_{Spec}$ and\\
 &  &  & [Fe/H]$_{Phot}$ is less $\leq$\ 0.5 dex and ``Spectroscopic" is used if only [Fe/H]$_{Spec}$\\
 &  &  & is available and ``Photometric" is used if only [Fe/H]$_{Phot}$ is available, while a choice\\
 &  &  & is made between [Fe/H]$_{Spec}$ and [Fe/H]$_{Phot}$ if both are available and the difference\\
 &  &  & is $>$ 0.5 dex)\\
74 & $V_{\rm mag}$ Reference & $-$ & The Reference for the \textit{V} magnitude of the star\\
75 & Distance AGAMA & $-$ & The Reference for the distance used in AGAMA (StarHorse prioritized over BJ21 unless StarHorse\\
 &  &  & distance has relative error greater than 0.3, if both have a relative error greater than 0.3\\
 &  &  & we adopt no distance estimate)\\
76 & RV Reference & $-$ & The Reference for the RV\\
77 & (v$_{\rm r}$,v$_{\phi}$,v$_{\rm z}$)  & (km s$^{-1}$) & The cylindrical velocities of the star as given by AGAMA\\
78 & Error & (km s$^{-1}$) & The cylindrical velocity errors of the star as given by Monte Carlo sampling through AGAMA\\
79 & (J$_{\rm r}$,J$_{\phi}$,J$_{\rm z}$) & (kpc km s$^{-1}$) & The cylindrical actions of the star as given by AGAMA\\
80 & Error & (kpc km s$^{-1}$) & The cylindrical action errors of the star as given by Monte Carlo sampling through AGAMA\\
81 & Energy & (km$^{2}$ s$^{-2}$) & The orbital energy of the star as given by AGAMA\\
\pagebreak \\[-3.5ex]
82 & Error  & (km$^{2}$ s$^{-2}$) & The orbital energy error of the star as given by Monte Carlo sampling through AGAMA\\
83 & $r_{\rm peri}$ & (kpc) & The Galactic pericentric distance of the star as given by AGAMA\\
84 & Error & (kpc) & The Galactic pericentric distance error of the star as given by Monte Carlo sampling through\\
 &  &  & AGAMA\\
85 & $r_{\rm apo}$ & (kpc) & The Galactic apocentric distance of the star as given by AGAMA\\
86 & Error & (kpc) & The Galactic apocentric distance error of the star as given by Monte Carlo sampling through\\
 &  &  & AGAMA\\
87 & Z$_{\rm max}$ & (kpc) & The maximum height above the Galactic plane of the star as given by AGAMA\\
88 & Error & (kpc) & The maximum height above the Galactic plane error of the star as given by Monte Carlo sampling\\
 &  &  & through AGAMA\\
89 & Eccentricity & $-$ & The eccentricity of the star given by ($r_{\rm{apo}}$ $-$ $r_{\rm{peri}}$)/($r_{\rm{apo}}$ $+$\\
 &  &  & $r_{\rm{peri}}$) through AGAMA\\
90 & Error & $-$ & The eccentricity error of the star as given by Monte Carlo sampling through AGAMA\\
\enddata
\end{deluxetable*}

\startlongtable\begin{deluxetable*}{l  l  c  c  c  c  c  c  c}
\tabletypesize{\scriptsize}
\tablecaption{Identified CEMP stars and their Group Association \label{tab:cemp}}
\tablehead{\colhead{Name} & \colhead{Group} & \colhead{T$_{\rm eff}$ (K)} & \colhead{log $g$} & \colhead{[Fe/H]} & \colhead{[C/Fe]} & \colhead{[C/Fe]$_{c}$} & \colhead{AC$_{c}$} & \colhead{[$\alpha$/Fe]}}
\startdata
J000022.60$-$130228.0&II&$4874$&$1.451$&$-2.936$&$+0.531$&$+0.951$&$6.445$&$+0.206$\\
J002330.69$-$163143.2&II&$5590$&$2.801$&$-2.474$&$+0.785$&$+0.795$&$6.751$&$+0.298$\\
J004539.31$-$745729.4&I&$5049$&$1.800$&$-2.242$&$+1.025$&$+1.185$&$7.373$&$-0.015$\\
J011522.35$-$802241.1&$\dots$&$4682$&$1.286$&$-2.484$&$+0.151$&$+0.701$&$6.647$&$+0.420$\\
J012931.10$-$160046.0&II&$5096$&$1.591$&$-2.707$&$+0.533$&$+0.853$&$6.576$&$+0.343$\\
J013346.60$-$272737.0&I&$4821$&$1.324$&$-2.787$&$+2.317$&$+2.447$&$8.090$&$+0.600$\\
J014738.05$-$113047.6&II&$4874$&$1.514$&$-2.897$&$+0.705$&$+1.075$&$6.608$&$\dots$\\
J023240.43$-$370811.2&II&$4784$&$0.871$&$-2.554$&$+0.088$&$+0.758$&$6.634$&$\dots$\\
J023558.70$-$674552.0&I&$4718$&$1.326$&$-2.121$&$+1.470$&$+1.600$&$7.909$&$+1.006$\\
J024010.80$-$141630.0&I&$5952$&$2.387$&$-2.068$&$+2.159$&$+2.179$&$8.541$&$+0.547$\\
J024620.10$-$151842.0&II&$4943$&$1.395$&$-2.882$&$+0.423$&$+0.893$&$6.441$&$+0.373$\\
J025007.20$-$514515.0&I&$4550$&$0.936$&$-3.015$&$+0.938$&$+1.518$&$6.933$&$\dots$\\
J032115.90$-$801824.0&I&$4550$&$1.081$&$-2.281$&$+0.475$&$+0.965$&$7.114$&$\dots$\\
J034236.50$-$133553.0&I/III&$6375$&$2.910$&$-3.222$&$+1.699$&$+1.709$&$6.917$&$\dots$\\
J040618.20$-$030525.0&I&$5024$&$1.444$&$-2.088$&$+0.545$&$+0.885$&$7.227$&$\dots$\\
J041716.50$-$033631.0&I&$4712$&$0.844$&$-2.891$&$+1.421$&$+1.821$&$7.360$&$+0.104$\\
J042314.50$+$013048.0&II&$5109$&$1.681$&$-2.821$&$+0.589$&$+0.829$&$6.438$&$+0.206$\\
J043831.80$-$705319.0&I&$5043$&$1.758$&$-2.355$&$+0.557$&$+0.797$&$6.872$&$\dots$\\
J044208.20$-$342114.0&I&$4676$&$0.811$&$-3.084$&$+2.096$&$+2.346$&$7.692$&$\dots$\\
J044240.00$-$104324.0&II&$6012$&$3.216$&$-2.597$&$+0.931$&$+0.941$&$6.774$&$\dots$\\
J045322.60$-$221040.0&I/II&$6322$&$3.265$&$-2.336$&$+0.980$&$+0.990$&$7.084$&$\dots$\\
J050947.56$-$442316.8&I&$5932$&$3.824$&$-2.218$&$+0.940$&$+0.940$&$7.152$&$+0.445$\\
J052029.30$-$582530.0&II&$4819$&$0.948$&$-3.023$&$+0.100$&$+0.820$&$6.227$&$\dots$\\
J055142.20$-$332734.0&II&$4964$&$1.247$&$-2.578$&$+0.284$&$+0.824$&$6.676$&$\dots$\\
J055736.90$-$512721.0&I&$4710$&$0.913$&$-3.015$&$+2.228$&$+2.408$&$7.823$&$\dots$\\
J060516.00$-$490728.0&I&$6923$&$3.250$&$-1.151$&$+1.109$&$+1.119$&$8.398$&$\dots$\\
J060554.00$-$330629.0&I&$5315$&$2.465$&$-2.155$&$+1.340$&$+1.360$&$7.635$&$+0.321$\\
J060555.10$-$485437.0&$\dots$&$5302$&$3.145$&$-2.382$&$+0.719$&$+0.729$&$6.777$&$+0.300$\\
J061950.00$-$531212.0&I&$5414$&$2.398$&$-2.107$&$+0.709$&$+0.719$&$7.042$&$\dots$\\
J065929.50$-$540613.0&I&$4794$&$1.240$&$-2.338$&$+0.619$&$+1.049$&$7.141$&$\dots$\\
J070520.28$-$334324.3&I/II&$5050$&$1.652$&$-2.367$&$+0.531$&$+0.841$&$6.904$&$\dots$\\
J085805.90$-$080917.0&II&$4775$&$0.975$&$-3.136$&$+0.339$&$+1.039$&$6.333$&$+0.525$\\
J091348.80$-$091909.0&I&$7215$&$2.288$&$-1.858$&$+0.789$&$+0.799$&$7.371$&$+0.600$\\
J092556.55$-$345037.3&I&$6676$&$4.054$&$-2.530$&$+1.480$&$+1.480$&$7.380$&$\dots$\\
J094921.83$-$161722.0&I&$4757$&$1.424$&$-2.793$&$+1.213$&$+1.553$&$7.190$&$+0.789$\\
J100800.70$-$100852.0&I/II&$4571$&$0.848$&$-2.747$&$+0.494$&$+1.094$&$6.777$&$\dots$\\
J101621.57$-$295952.1&I&$5060$&$1.545$&$-2.085$&$+0.453$&$+0.783$&$7.128$&$\dots$\\
J104410.96$-$035856.3&I&$6337$&$3.215$&$-2.223$&$+1.563$&$+1.573$&$7.780$&$+0.511$\\
J111912.33$-$160946.8&I&$4988$&$1.377$&$-2.844$&$+1.168$&$+1.528$&$7.114$&$\dots$\\
J112229.90$-$245854.0&I&$4761$&$2.574$&$-1.087$&$+0.930$&$+0.950$&$8.293$&$\dots$\\
J112243.40$-$020937.0&I/II&$6377$&$3.278$&$-2.748$&$+1.144$&$+1.144$&$6.826$&$\dots$\\
J114638.83$-$042239.5&I&$4326$&$0.433$&$-2.322$&$+0.177$&$+0.777$&$6.885$&$+0.274$\\
J114650.70$-$140643.0&III&$6434$&$2.033$&$-3.412$&$+1.616$&$+1.626$&$6.644$&$\dots$\\
J114948.00$-$081721.0&I&$5706$&$1.419$&$-3.035$&$+2.636$&$+2.746$&$8.141$&$+0.715$\\
J115337.30$-$020037.0&I/II&$5146$&$2.315$&$-2.593$&$+1.012$&$+1.032$&$6.869$&$+0.211$\\
J115801.30$-$152218.0&II&$5038$&$1.819$&$-2.953$&$+0.605$&$+0.745$&$6.222$&$\dots$\\
J121000.70$-$313036.0&II&$4914$&$1.195$&$-2.570$&$+0.273$&$+0.833$&$6.693$&$\dots$\\
J121920.34$-$172154.6&II&$5070$&$1.815$&$-2.657$&$+0.536$&$+0.706$&$6.479$&$+0.365$\\
J123550.13$-$313111.2&II&$5123$&$2.224$&$-2.428$&$+0.789$&$+0.799$&$6.801$&$+0.272$\\
J124537.80$-$092704.0&I&$5450$&$3.288$&$-1.744$&$+0.817$&$+0.827$&$7.513$&$+0.478$\\
J124718.30$-$555047.0&I&$4917$&$1.678$&$-2.090$&$+1.347$&$+1.467$&$7.807$&$+0.052$\\
J125728.30$-$512522.0&$\dots$&$4985$&$1.513$&$-2.356$&$+0.294$&$+0.724$&$6.798$&$\dots$\\
J130119.30$-$454552.0&$\dots$&$4769$&$0.767$&$-2.417$&$+0.050$&$+0.710$&$6.723$&$\dots$\\
J130744.20$-$154542.0&I/II&$6210$&$3.035$&$-2.405$&$+0.820$&$+0.830$&$6.855$&$+0.257$\\
J132418.00$-$280356.0&I/II&$4757$&$1.167$&$-2.589$&$+0.717$&$+1.197$&$7.038$&$\dots$\\
J132918.40$-$025739.0&I&$7147$&$3.168$&$-1.269$&$+1.100$&$+1.120$&$8.281$&$+0.600$\\
J133245.30$-$240326.0&II&$4589$&$0.419$&$-2.440$&$+0.087$&$+0.747$&$6.737$&$\dots$\\
J134603.40$-$052414.0&II&$5372$&$2.880$&$-2.831$&$+0.855$&$+0.865$&$6.464$&$+0.535$\\
J135439.20$-$412746.0&I&$5072$&$1.745$&$-2.278$&$+0.546$&$+0.786$&$6.938$&$\dots$\\
J140230.10$-$053905.0&II/III&$6504$&$3.582$&$-3.067$&$+1.259$&$+1.259$&$6.622$&$+0.187$\\
J141126.20$-$113746.0&II&$4301$&$0.658$&$-3.308$&$+0.237$&$+0.977$&$6.099$&$+0.546$\\
J143811.20$-$212009.0&II&$4542$&$1.186$&$-2.860$&$+0.556$&$+1.106$&$6.676$&$\dots$\\
J143926.81$-$194913.3&I&$5000$&$1.512$&$-2.830$&$+1.365$&$+1.645$&$7.245$&$\dots$\\
J151335.50$-$124434.4&II&$5097$&$1.776$&$-2.846$&$+0.593$&$+0.743$&$6.327$&$+0.144$\\
J151919.20$-$083234.0&I/II&$6507$&$4.032$&$-2.364$&$+0.899$&$+0.899$&$6.965$&$+0.182$\\
J152140.00$-$353810.0&II&$6029$&$3.024$&$-2.955$&$+1.151$&$+1.161$&$6.636$&$+0.928$\\
J152713.50$-$233618.0&I&$6255$&$4.199$&$-1.973$&$+0.765$&$+0.765$&$7.222$&$+0.297$\\
J155751.80$-$083955.0&II&$4572$&$0.699$&$-2.684$&$+0.377$&$+0.987$&$6.733$&$\dots$\\
J155846.99$-$033950.8&I/II&$5216$&$1.994$&$-2.685$&$+1.224$&$+1.264$&$7.009$&$\dots$\\
J160334.40$-$062337.0&III&$4548$&$1.494$&$-3.314$&$+1.260$&$+1.650$&$6.766$&$\dots$\\
J161156.70$-$803442.0&I/II&$4467$&$0.842$&$-2.419$&$+0.619$&$+1.109$&$7.120$&$+0.437$\\
J162051.23$-$091258.0&II&$4908$&$1.589$&$-3.103$&$+0.415$&$+0.755$&$6.082$&$+0.325$\\
J165143.93$-$202926.5&$\dots$&$6388$&$2.245$&$-2.360$&$+0.740$&$+0.750$&$6.820$&$\dots$\\
J171800.40$-$033535.0&I&$7766$&$4.731$&$-1.492$&$+0.933$&$+0.933$&$7.871$&$\dots$\\
J171813.40$-$682301.0&II&$4926$&$1.364$&$-2.563$&$+0.200$&$+0.720$&$6.587$&$\dots$\\
J173600.10$-$514530.0&$\dots$&$4862$&$1.012$&$-2.356$&$+0.109$&$+0.749$&$6.823$&$\dots$\\
J180242.30$-$440443.0&II&$4463$&$0.732$&$-2.749$&$+0.279$&$+0.919$&$6.600$&$\dots$\\
J190733.10$-$415832.0&II&$4602$&$1.110$&$-2.753$&$+0.088$&$+0.758$&$6.435$&$\dots$\\
J192325.19$-$583340.9&I&$5056$&$1.661$&$-2.396$&$+0.984$&$+1.214$&$7.248$&$\dots$\\
J193204.50$-$730918.0&II&$4300$&$0.308$&$-3.683$&$+0.674$&$+1.404$&$6.151$&$\dots$\\
J194318.70$-$594819.9&I&$5225$&$1.882$&$-2.228$&$+0.741$&$+0.861$&$7.063$&$-0.095$\\
J194411.10$-$111927.0&I&$5418$&$0.784$&$-2.185$&$+0.172$&$+0.752$&$6.997$&$+0.814$\\
J194813.29$-$490344.8&II&$4830$&$1.090$&$-2.486$&$+0.135$&$+0.765$&$6.709$&$\dots$\\
J195637.98$-$793124.0&I&$6150$&$3.545$&$-0.930$&$+1.041$&$+1.041$&$8.541$&$+0.439$\\
J200630.40$-$490004.0&I&$6368$&$4.370$&$-1.611$&$+0.887$&$+0.887$&$7.706$&$+0.233$\\
J201446.10$-$563530.0&I&$4785$&$1.334$&$-2.849$&$+1.387$&$+1.707$&$7.288$&$+0.547$\\
J202434.50$-$423334.0&II&$5394$&$3.155$&$-2.633$&$+0.901$&$+0.911$&$6.708$&$+0.469$\\
J202602.40$-$474029.0&II&$5041$&$1.650$&$-2.542$&$+0.475$&$+0.785$&$6.673$&$\dots$\\
J203622.60$-$071420.0&II&$4191$&$0.203$&$-2.882$&$+0.236$&$+0.916$&$6.464$&$+0.212$\\
J203706.40$-$122125.0&II&$4562$&$1.048$&$-2.860$&$+0.334$&$+0.974$&$6.544$&$+0.141$\\
J203740.42$-$552050.0&II&$4912$&$1.008$&$-2.811$&$+0.476$&$+1.086$&$6.705$&$\dots$\\
J203843.20$-$002333.0&II&$4746$&$0.735$&$-3.263$&$-0.035$&$+0.735$&$5.902$&$+0.426$\\
J204406.30$-$300007.0&I&$6059$&$3.350$&$-2.286$&$+0.788$&$+0.788$&$6.932$&$-0.145$\\
J204450.70$-$371400.0&II&$4507$&$0.717$&$-3.023$&$+0.341$&$+1.021$&$6.428$&$+0.266$\\
J210540.70$-$452057.0&II&$5074$&$1.778$&$-2.637$&$+0.601$&$+0.771$&$6.564$&$\dots$\\
J210634.70$-$495750.0&II&$5849$&$3.366$&$-2.549$&$+0.839$&$+0.839$&$6.720$&$\dots$\\
J210642.94$-$682826.7&II&$5401$&$3.696$&$-2.918$&$+1.019$&$+1.019$&$6.531$&$+0.659$\\
J210958.10$-$094540.0&II&$4625$&$0.542$&$-3.129$&$-0.049$&$+0.711$&$6.012$&$+0.501$\\
J211525.50$-$150331.0&I/II&$4622$&$0.728$&$-2.601$&$+0.504$&$+1.074$&$6.903$&$+0.818$\\
J214055.30$-$101223.0&I&$7782$&$3.010$&$-0.988$&$+1.529$&$+1.549$&$8.991$&$\dots$\\
J222203.45$-$802459.3&II&$4799$&$1.500$&$-2.895$&$+0.442$&$+0.842$&$6.377$&$+0.236$\\
J222236.00$-$013827.1&II/III&$5197$&$2.377$&$-3.049$&$+1.251$&$+1.261$&$6.642$&$+0.286$\\
J223733.20$-$434118.0&II&$5079$&$1.714$&$-2.406$&$+0.511$&$+0.781$&$6.805$&$\dots$\\
J224601.31$-$593804.5&II&$4584$&$0.905$&$-2.543$&$+0.152$&$+0.802$&$6.689$&$+0.483$\\
J225319.90$-$224856.0&I&$4864$&$1.167$&$-2.312$&$+1.402$&$+1.572$&$7.690$&$+0.465$\\
J233107.20$-$022330.0&I&$4867$&$1.423$&$-3.292$&$+2.449$&$+2.609$&$7.747$&$+0.600$\\
\enddata
\tablecomments{The CEMP Groups are given in \citet{Yoon2016}.}
\end{deluxetable*}

\bibliography{main}{}
\bibliographystyle{aasjournal}

\end{document}